\documentclass[preprint,11pt]{JHEP3} % 10pt is ignored!

\JHEPspecialurl{http://jhep.sissa.it/JOURNAL/JHEP3.tar.gz}

\usepackage{epsfig,multicol,amsmath}

\usepackage{epstopdf}
\usepackage{rotating}

\def\beq{\begin{equation}}
\def\eeq{\end{equation}}
\def\bea{\begin{eqnarray}}
\def\eea{\end{eqnarray}}

\def\eq#1{{Eq.~(\ref{#1})}}
\def\fig#1{{Fig.~\ref{#1}}}
\newcommand{\bas}{\bar{\alpha}_S}
\newcommand{\as}{\alpha_S}

\newcommand{\Lb}{\left(}
\newcommand{\Rb}{\right)}

\parskip=\itemsep               %?

\def\thefootnote{\fnsymbol{footnote}}

\title{ The BFKL Pomeron Calculus in zero transverse dimensions: diffractive
processes
and survival  probability for  central diffractive  production  }
\author{\Large  M.
Kozlov ${}^{a)}$ \thanks{Email:
kozlov@post.tau.ac.il;}\,,\, E. Levin ${}^{a)}$ \thanks{Email:
leving@post.tau.ac.il,
levin@mail.desy.de;} \,,\,V. Khachatryan ${}^{b)}$
\thanks{Email:khavladi@yerphi.am;}\,\,\,\,\,\,and\,\,\,J.
Miller ${}^{a)}$ \thanks{Email:jeremymiller@london.com,jeremy@post.tau.ac.il;}
\\
a) Department of Particle Physics, School of Physics and Astronomy\\ Raymond and
Beverly Sackler Faculty
of Exact Science\\  Tel Aviv University, Tel Aviv, 69978, Israel\\
b) Theory Division, Yerevan Physics Institute, Yerevan, 375036, Armenia}

\abstract{In this paper we discuss the processes of diffractive production in
the framework of the BFKL Pomeron calculus in
zero transverse dimension. Considering the diffractive production of  a bunch of
particles with not very large masses,
namely,
 $\ln\Lb  M^2/m^2 \Rb\,\,\ll \,\,\frac{1}{\bas}\,\ln\Lb
\frac{N^2_c}{\bas^2}\Rb$,
we found explicit formulae for calculation of the cross sections for the single
and double
 diffractive production as
 well as for the value of the survival probability for the diffractive central
production.   These formulae include the
 influence of
the correlations due to so called Pomeron loops  on the values of all discussed
observables.  The comparison with the
other approaches on
the market is given. The main conclusion of this
comparison: the 
Mueller-Patel-Salam-Iancu approximation gives sufficiently good 
descriptions and close to the exact result for elastic and diffractive 
cross section but considerable overshoot the value of the survival 
probability.  }

 \keywords{BFKL Pomeron,  Diffractive cross sections,  Mean field approach, 
Pomeron loops, Exact solution }
\dedicated{PACS: 13.85-t, 13.85.Hd, 11.55.-m, 11.55.Bq}
\preprint{TAUP - 2839/06 \\
hep-ph/0610084 \\
\today}
\begin{document}

\def\thefootnote{\arabic{footnote}}
\section{Introduction}
\label{sec:Int}
Diffractive production processes, being pure quantum mechanical by their origin,
have always been a source of
interesting
information and sometimes even of confusion for all theoretical approaches to
high energy  scattering. It is enough to
 mention that due to the AGK cutting rules \cite{AGK} the total  cross section
of the diffractive production
is equal to the contribution of the first shadowing correction (with opposite
sign)  to the total cross section.

The goal of this paper is to discuss these processes in the framework of high
density QCD \cite{GLR,MUQI,MV}.
Since the most simple and elegant approach, that allows us to discuss both the
total cross sections and the diffractive production on the same footing,  is
the AGK cutting rules,  we choose the particular version of the theoretical
approach
to high density QCD: the BFKL  Pomeron calculus  \cite{GLR,MUQI}. This approach
is based on the BFKL Pomeron
\cite{BFKL} and the  interactions between the BFKL Pomerons
\cite{BART,BRN,NP,BLV}, which are
taken into account in the spirit of the  Gribov Reggeon Calculus \cite{GRC}. 
The BFKL Pomeron calculus
can be formulated  at least in two beautiful forms:  as  the functional integral
\cite{BRN} and as the evolution equation for
 the generating functional (see Refs.\cite{MUCD,L1,L2,L3,L4}) . Both of them 
reproduce the mean field approximation for
the total cross sections (see Refs. \cite{MV,BK,JIMWLK}) as well as leading to
the mean field equations for the diffractive
 amplitude (see Ref. \cite{KL}). In both of them we can include so called
Pomeron
(see Refs.\cite{IT,MSW,L3,IST,KOLU,HIMS})
loops which were not
taken into account in the mean field approximation and which are very essential
for the processes of the diffraction
 production especially in the region of large masses of the diffractive system.
However, it is not clear at the moment the interrelation between the BFKL
Pomeron calculus and the  first attempt to
 suggest the equations for the diffractive processes in the framework of the
approach to high density is more
 general than the mean field approach (see Ref. \cite{HIMST}).

In this paper we discuss the diffractive production in the simplest version of
the BFKL Pomeron calculus, namely, in the
toy model in which we neglect the fact  that the QCD interaction can change the
sizes of interacting dipoles.  Such system of
 colourless dipoles reduces to the BFKL Pomeron calculus in zero transverse
dimension.  The analytical solution for the
 generating functional has been found in this model (see Ref. \cite{KLTM} and an
earlier attempt to solve this problem in
 Refs. \cite{L4,REST,SHXI}).  Using this solution as well as the AGK cutting
rules we can develop the theoretical
 approach to the diffractive processes. The first attempt to build such an
approach has been made in Ref. \cite{SHBOW},
 however it was done in the limited kinematic region.

The paper is organized as follows. In the next section we discuss the single
diffractive production in  both the mean
field approximation and in the approximation that includes the Pomeron loops. 
The equations that govern this process, are
 written and the analytical solutions are found. However, we are able to solve
the problem only in the restricted range of
 mass for the diffractive system ($M$), namely,
\beq \label{MR}
\frac{1}{\bas}\,\,\,\,\ll\,\,\,\,Y_0\,\,\equiv\,\,\ln\Lb  M^2/m^2 \Rb\,\,\ll
\,\,
\frac{1}{\bas}\,\ln\Lb \frac{N^2_c}{\bas^2} \Rb
\eeq
where $m$ is a typical mass of soft interaction (say $m = 1 \mbox{GeV}$) and   
$\as$ is the running QCD coupling $\bas\,\,=\,\,(N_c/\pi)\as$.
 In this region we can neglect the fluctuations in the diffractively produced
cascade of partons
due to the
  Pomeron loops contribution \cite{MUPA,IM,KOLE,KODD,SHBOW}.  It should be
stressed that the Pomeron loops are taken
into account  for the ``elastic" scattering with rapidity
  \beq \label{MR1}
  Y\,\,-\,\,Y_0\,\,=\,\,\ln\Lb s/M^2 \Rb\,\,\geq\,\,\frac{1}{\bas}\,\ln\Lb
\frac{N^2_c}{\bas^2} \Rb
\eeq

  In the third section we consider the double diffractive dissociation in the
kinematic region of \eq{MR} for both
 produced masses ($M_1$ and $M_2$)  but at very high energy
  \beq \label{MR2}
\ln\Lb \frac{s\,m^2}{M_1^2\,M^2_2} \Rb\,\,\,\geq\,\,\frac{1}{\bas}\,\ln\Lb
\frac{N^2_c}{\bas^2} \Rb\,\,.
\eeq
The fourth section is devoted to calculations of the survival probability for a
process with a large rapidity
gap (LRG)  of the following type
\bea
\mbox{hadron}\,\,+\,\,\mbox{hadron}\,\,&\longrightarrow& \label{MR3}\\
\mbox{hadron}\,\,\,+\,\,\,\,\mbox{\{ LRG \}}\,\,&+&\,\,\mbox{\{system of fixed
mass ( di-jet,
Higgs meson ,...)\}}\,\,+
\,\,\mbox{\{ LRG \}}\,\,+\,\,\mbox{hadron} \nonumber
\eea
which we can call a central diffractive process. The value of the survival 
probability of this process is a hot question
 since this reaction is one of the best for measuring Higgs meson at the LHC (
see Refs. \cite{DG,DGK,GLMSB,BBKM,DGHP,JM}).

In section five we compare our estimates with other approaches that are on the
market and in conclusions we
summarize the results.

Our approach in this paper is very close to one that has been developed in Ref.
\cite{SHBOW}.  Therefore, it is worthwhile to
stress the points on which we differ:
\begin{enumerate}
\item \quad  In both papers the range of produced mass is restricted by \eq{MR},
and, therefore,  we can neglect the
fluctuations due to the Pomeron loops in produced system.  It results in using
the mean field approximations (`fan', tree
Pomeron diagrams) for the produced system in both papers. However, we did not
assume that the total energies restricted
 ( $Y  - Y_0 \,\ll\,\frac{1}{\bas}\,\ln\Lb \frac{N^2_c}{\bas^2} \Rb$) as it is
done in Ref. \cite{SHBOW} to justify the
 use of Mueller-Patel-Salam-Iancu approach \cite{MUPA,MS,IM}. For region, given
by \eq{MR1}, we use the exact solution
 of the
BFKL Pomeron calculus that has been found in Ref.\cite{KLTM};
\item \quad  In our approach we explore the Kovchegov-Levin  equation \cite{KL} (see also Refs. \cite{WE,KLW} where this equation was generalized in the mean field approximation) 
for diffractive production in the mean
field approximation. Being equivalent to AGK cutting rules  in the BFKL Pomeron
calculus in zero transverse dimension,
it has solid
foundation in high density QCD (see Refs. \cite{KL,HIMST}) and allows us to
develop a more general formalism in calculating
 the diffractive observables. In this  formalism the solution to Kovchegov-Levin
equation for $Y_0 $ in the kinematic range
of \eq{MR} plays role of the initial condition for the general equation for the
scattering amplitude in the rapidity range
 of $Y -Y_0$;
\item \quad  Even in section 5.2 where we apply our approach to the
Mueller-Patel-Salam-Iancu range of energies we obtain
more general formulae since we do not assume that $Y -Y_0 \,\gg\,Y_0$ as in Ref.
\cite{SHBOW}.
\end{enumerate}

One of our motivations to discuss  diffractive processes stems from the
interesting paper of
 Y.~Hatta, E.~Iancu, C.~Marquet, G.~Soyez and D.~N.~Triantafyllopoulos
\cite{HIMST} in which the diffractive dissociation
 has been considered in high density QCD. In this  paper a generalization of
Kovchegov-Levin equation \cite{KL}
 is proposed.
We wished to check and to confront these equations with the simplest model that
we have at our disposal: the BFKL Pomeron
 calculus in zero transverse dimension.  Indeed, we found two features of the
diffractive processes in our approach
 that support the main ideas of Ref. \cite{HIMST}: (i) a kind of factorization
between wee partons  (Pomerons)
 produced by the
 incoming particle at rapidity $Y_0$ and the process of creation of the
diffractive system by these partons  (Pomerons);
and (ii) an impressive difference between the mean field approximation and the
exact answer.

\newpage

\section{Single diffractive production}
\label{sec:SD}
\subsection{The mean field approximation.}
\subsubsection{General approach.}

\FIGURE[h]{
\begin{minipage}{65mm}{
\centerline{\epsfig{file=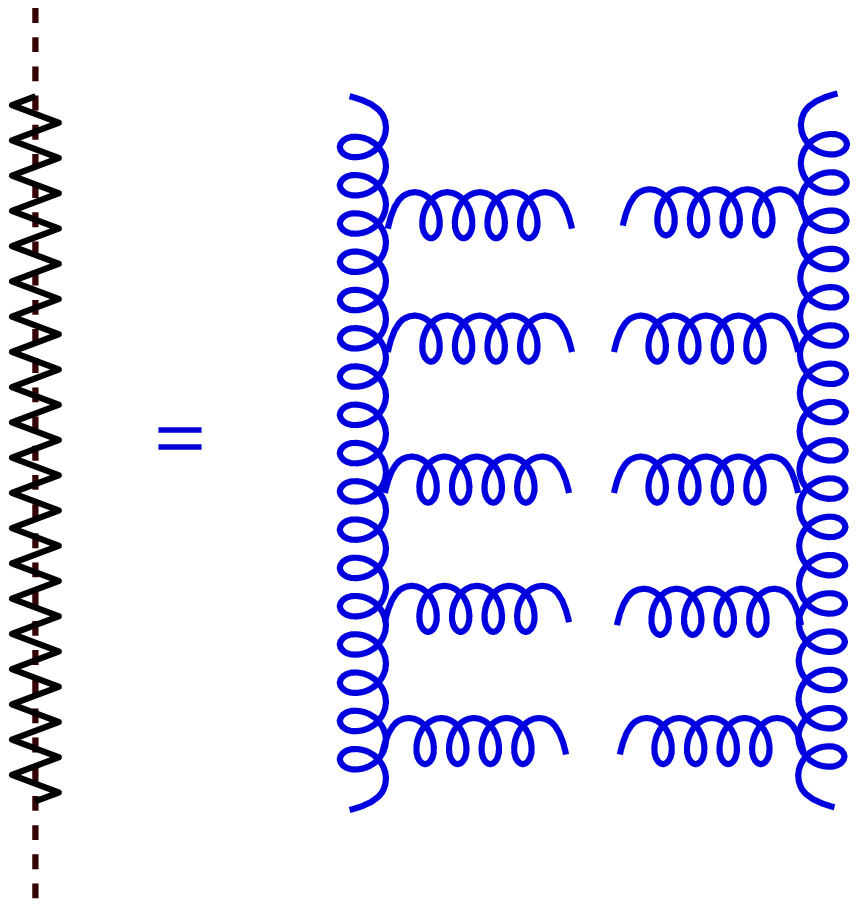,width=60mm}}
}
\end{minipage}
\caption{The diagram for the cut BFKL Pomeron that illustrates \eq{UCP}.  }
\label{cutp}
}

Our approach to diffractive production is based on the AGK cutting rules
\cite{AGK}. These rules use the unitarity
 constraints
 $s$-channel,
namely,
\beq \label{UCP}
2\,Im\,A^{BFKL}(s,b)\,\,=\,\,2\,\,N^{BFKL}(s,b)\,\,=\,\,G^{BFKL}_{in}(s,b)
\eeq
where $Im\,A^{BFKL}(s,b) \equiv\,N^{BFKL}(s,b)$ denotes the imaginary part of
the  elastic scattering amplitude
for dipole-dipole
interaction at energy $W= \sqrt{s}$ and at the impact parameter $b$. It is
normalized in a way that  the total cross
section is equal
 to
 $\sigma_{tot}\,\,=\,\,2\,\int \,d^2b\,\,N^{BFKL}(s,b)$.
 $G^{BFKL}_{in}$ is the contribution of all inelastic  processes for the BFKL
Pomeron and $\sigma_{in} =\int \,d^2b\,
\,G_{in}(s,b)$.
 Therefore, \eq{UCP} gives us the structure of the BFKL Pomeron exchange through
the inelastic processes and it can be
 formulated as
 the statement that the exchange of the BFKL Pomeron is related to the processes
of multi-gluon production in a certain
 kinematics
(see \fig{cutp}). This equation was proven in Ref. \cite{BFKL}.  Using \eq{UCP},
we can express all processes and,
 in particular,
 the processes of diffractive production in terms of exchange and interactions 
of
 the BFKL Pomerons and/or the cut BFKL Pomerons (see \fig{cutps}).

\FIGURE{
\centerline{\epsfig{file=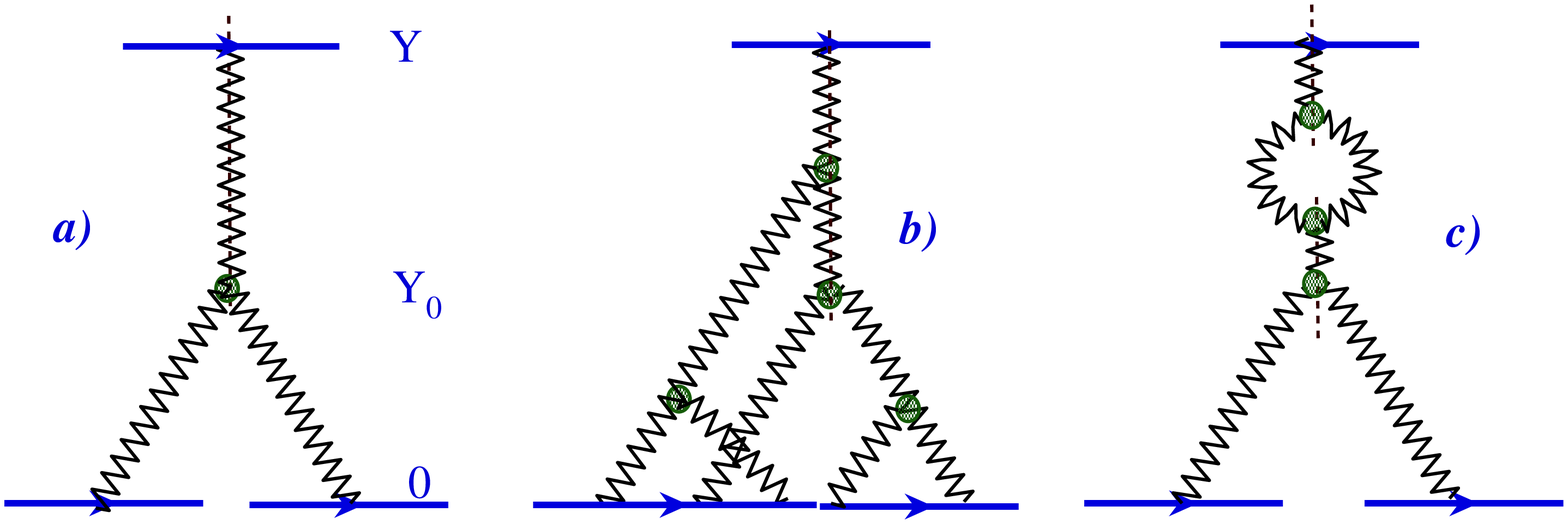,width=160mm,height=30mm}}
\caption{ Several examples of the Pomeron diagrams that contribute to the
diffractive production of the bunch of
  gluon with
$\ln(M^2/m^2)\,=\,Y - Y_0$ : diagram $a)$ is the first diagram for the  process
of the difractive production, while
diagram $b)$ gives
an example of more complicated contribution to the diffractive production in the
mean field approximation;   diagram $c)$ is the example of the
diagrams that
 we neglected in the mean field approximation as well as in the approach of this
paper. }
\label{cutps}
}

The AGK cutting rules establish the relation between different processes that
stem from   BFKL Pomeron diagrams. For
 example,
 the simple triple Pomeron diagrams in \fig{agk3p}     leads to three inelastic
processes: the diffractive production of
 the system with
 mass $\ln(M^2/m^2)\,=\,Y - Y_0$ (\fig{agk3p} -A); the multi-gluon production in
the entire kinematic region of
rapidity $Y - 0$
with the same multiplicity of gluons as in one Pomeron ( \fig{agk3p} -B); and
the   multi-gluon production in the
 region $Y -0$ but with
 the same multiplicity of gluons as in one Pomeron only in the rapidity window
$Y - Y_0$ while for the rapidity
$Y_0 - 0$ the gluon
 multiplicity in two times larger than for one Pomeron( \fig{agk3p} -C).
The AGK cutting rules \cite{AGK} say that the cross sections of these three
processes are related as
\beq \label{AGK3P}
\sigma_A \,\,:\,\,\sigma_B
\,\,:\,\,\sigma_C\,\,\,\,\,=\,\,\,\,2\,\,:\,\,-4\,\,:\,\,1
\eeq
At first sight the cross section of the process B is negative but it should be
stressed that one Pomeron also
contributes to the same
process and the resulting contribution is always positive.

\FIGURE{
\centerline{\epsfig{file=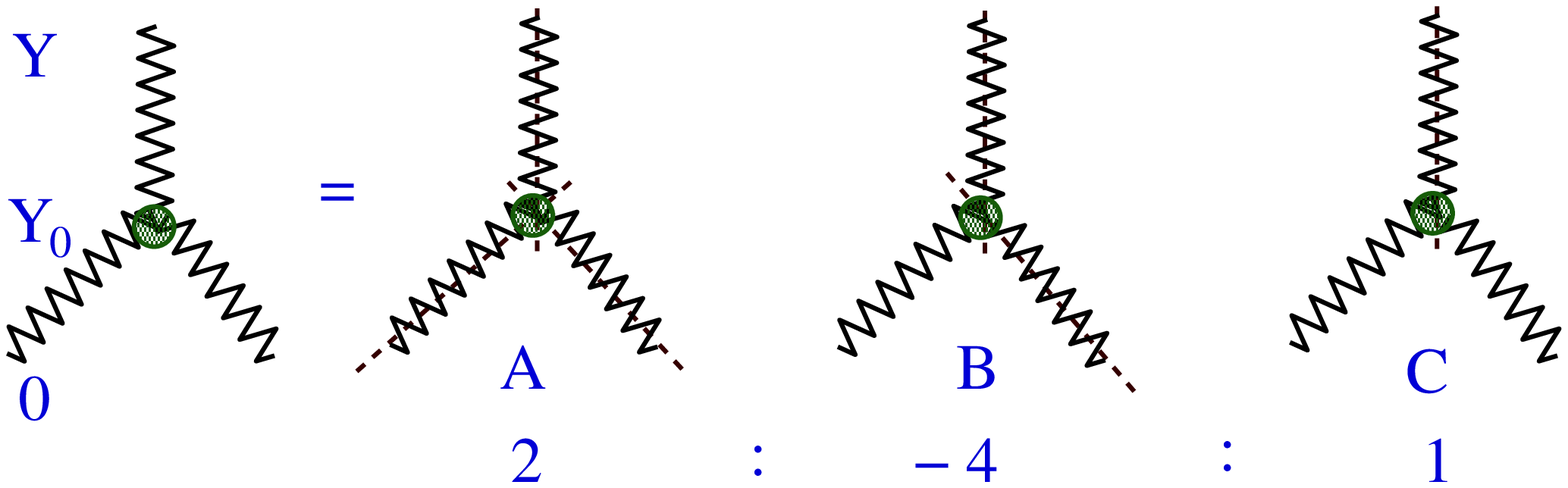,width=140mm}}
\caption{ The AGK cutting rules for the triple Pomeron diagram: the diffractive
production of the system with mass
 $\ln(M^2/m^2)\,=\,Y - Y_0$ (\fig{agk3p} -A); the multi-gluon production in the
entire kinematic region of
 rapidity $Y - 0$
 with the same multiplicity of gluons as in one Pomeron ( \fig{agk3p} -B); and
the   multi-gluon production in
the region $Y -0$
 but with the same multiplicity of gluons as in one Pomeron only in the rapidity
window $Y - Y_0$ while for the
rapidity $Y_0 - 0$
the gluon  multiplicity in two times larger than for one Pomeron( \fig{agk3p}
-C). The zigzag line denotes the Pomeron exchange while the zigzag line crossed
y the dotted line stands for the cut Pomeron. }
\label{agk3p}
}

\fig{cutp}, \fig{cutps} and  \fig{agk3p} as well as \eq{AGK3P} allow us to
understand
the equation for the single diffractive production in the mean field
approximation that  has been written by
Kovchegov and Levin
\cite{KL} and that has the following form (see \fig{kleq} for graphical
representation of this equation)
\FIGURE{
\centerline{\epsfig{file=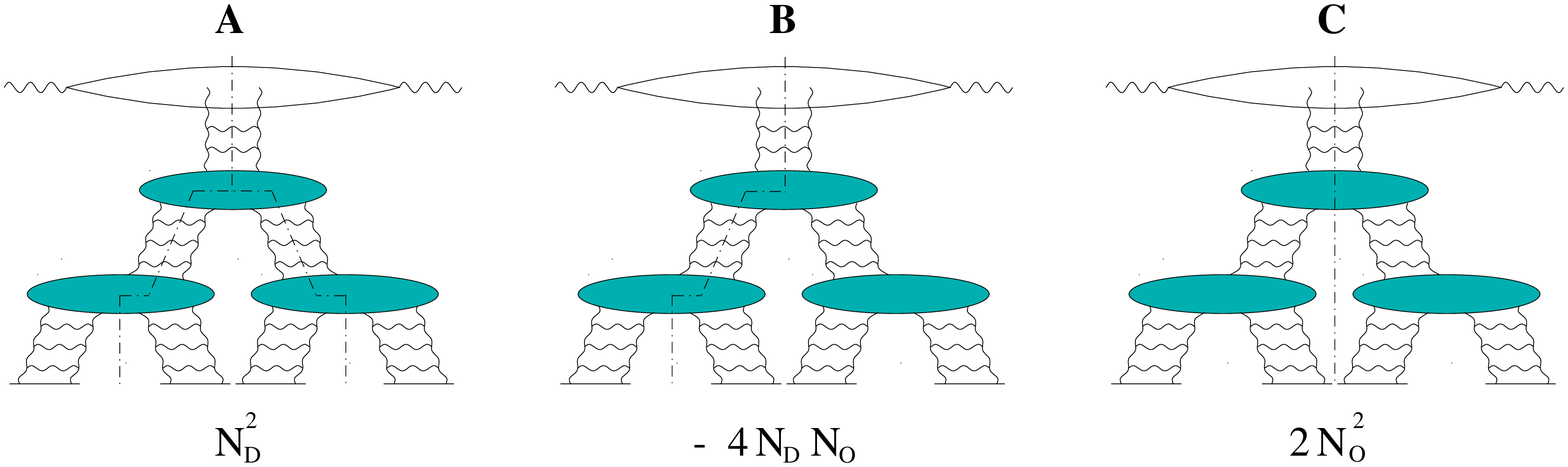,width=140mm}}
\caption{ Different Pomeron cuts  contributing to the cross section
of diffractive dissociation which lead to different terms on the right
hand side in \eq{KLND}}
\label{kleq}
}

\beq \label{KLND}
  \frac{\partial N^D ({\bf x}_{01},{\bf b}, Y, Y_0)}{\partial Y} =
  \frac{ \alpha C_F}{\pi^2} \, \int_\rho d^2 x_2 \,\,
  \frac{x^2_{01}}{x^2_{02} x^2_{12}}
\eeq
\bea
&&\left(\, N^D ({\bf x}_{02},{\bf b} + \frac{1}{2} {\bf x}_{12}, Y,
Y_0)\,\,+\,\,N^D ({\bf
x}_{12},{\bf b} + \frac{1}{2} {\bf x}_{02}, Y, Y_0)\,\,-\,\,N^D ({\bf
x}_{01},{\bf b}, Y,
Y_0)    \right. \nonumber\\
&& \left. +\, N^D ({\bf x}_{02},{\bf b} +
  \frac{1}{2} {\bf x}_{12}, Y, Y_0) \, N^D ({\bf x}_{12},
  \frac{1}{2} {\bf x}_{02}, Y, Y_0)  \,\,\,\, \,\,\,\,\,  \,\,\,\, \,\,\,\,\,
\,\,\,\, \,\,\,\,\, \,\,\,\, \,\,\,\,\,\ \,\,\,\, \,\,\,\,\,
 \,\,\,\, \,\,\,\,\,  \,\,\,\, \,\,\,\,\, \,\,\,\, \,\,\,\,\, \,\,\,\,
\,\,\,\,\,\ \,\,\,\, \,\,\,\,\,  \mbox{(A)}
\right. \nonumber \\
&&\left.- 4 \, N^D ({\bf x}_{02},{\bf b} + \frac{1}{2} {\bf x}_{12}, Y, Y_0)
\, N_0 ({\bf x}_{12},{\bf b} + \frac{1}{2} {\bf x}_{02}, Y) \,\,\,\, \,\,\,\,\, 
\,\,\,\, \,\,\,\,\, \,\,\,\, \,\,\,\,\, \,\,\,\, \,\,\,\,\,\ \,\,\,\,
\,\,\,\,\,
 \,\,\,\, \,\,\,\,\,  \,\,\,\, \,\,\,\,\, \,\,\,\, \,\,\,\,\, \,\,\,\,
\,\,\,\,\,\ \,\,\,\, \,
\mbox{(B)} \right. \nonumber \\
&&\left. + 2 \, N_0 ({\bf x}_{02},{\bf b} + \frac{1}{2} {\bf x}_{12}, Y) \, N_0
({\bf x}_{12}, {\bf b} + \frac{1}{2} {\bf x}_{02}, Y) \right) \,\,\,\,
\,\,\,\,\,  \,\,\,\, \,\,\,\,\, \,\,\,\, \,\,\,\,\, \,\,\,\, \,\,\,\,\,\
\,\,\,\, \,\,\,\,\,
 \,\,\,\, \,\,\,\,\,  \,\,\,\, \,\,\,\,\, \,\,\,\, \,\,\,\,\, \,\,\,\,
\,\,\,\,\,\ \,\,\,\, \,\,\,\,\,\,
\mbox{(C)} \nonumber
\eea

The physical meaning of $N_D$ is the cross section of the diffractive produced
masses $M$ large than
$ \ln(M^/m^2)\,\geq \,Y - Y_0$ or in other words
\beq \label{DEF}
 N^D \Lb{\bf x}_{01},{\bf b}, Y, Y_0 \Rb\,\,\,=\,\,\int^Y_{Y_0}
\,\,d\,Y'\,\,N^{SD}\Lb{\bf x}_{01},{\bf b}, Y, Y' \Rb
 \eeq
 where $N^{SD}\Lb{\bf x}_{01},{\bf b}, Y, Y' \Rb$ is the cross section of the
diffractive production at energy $Y$  with
the produced mass equal to $\ln(M^2/m^2)\,=\,Y - Y'$.
 The initial condition for \eq{KLND} follows from the definition of  $N^D$ and
it has the form
with the initial condition given by
\beq \label{INCSD}
N^D_{ic}({\bf x}_\perp,{\bf b},  Y_0)\,\,=\,\,N^D ({\bf x}_\perp,{\bf b}, Y=Y_0,
Y_0) = 0\,; \,\,\mbox{with}\,\,
-\,\frac{\partial N^D ({\bf x}_\perp,{\bf b}, Y, Y_0) }{\partial \,Y_0}|_{Y =
Y_0}\,\,=\,\,
N_0^2 ({\bf x}_\perp,{\bf
b}, Y_0).
\eeq
where $N_0$ is the solution of the Balitsky - Kovhegov equation \cite{BK} and
$N_D(\bf{x},\bf{b};Y;Y_0)$ is the
diffraction dissociation of the colourless dipole with size $x_{\perp}$  at
impact parameter $b$ into the system of gluons with
 the rapidity gap larger than $Y_0$ at energy $Y$. At first sight, \eq{KLND} 
contradicts the AGK relations given by
\eq{AGK3P} (see \fig{agk3p}), but this  contradiction can be easily resolved if
we take into account coefficient 2
 in \eq{UCP} (see Refs. \cite{KL} for more details as well as for a proof based
directly on the dipole approach
 to high energy scattering).

Despite the simple structure of \eq{KLND} which is only a little bit more
complicated than the Balitsky - Kovhegov equation,
as far as we know there exists no analytical solution  to this equation and
there is the  only attempt to solve it
 numerically
 \cite{NSD}.  However, this equation has a simple solution in the toy model (see
Refs. \cite{KL,BGLM,DGBK} )   which we
 are going to discuss.
\subsubsection{The BFKL Pomeron calculus in zero transverse dimension.}
The mean field approximation looks extremely simple  in the BFKL Pomeron
calculus in zero transverse dimension
(the toy model
\cite{MUCD,L1,L2}) . In this model \eq{KLND} has a simple form
\beq \label{MF1}
\frac{d N_D\Lb u;Y,Y_0 \Rb }{d\, \Gamma(1 \to 2)\,Y}\,\,=\,\,N_D\Lb u;Y,Y_0 \Rb
\,\,+\,\,N^2_D\Lb u;Y,Y_0 \Rb \,\, - \,\,4\,N_D\Lb u;Y,Y_0 \Rb \,N_0\Lb u;Y
\Rb\,\,\,+\,\,\,2\,N^2_0\Lb u;Y \Rb
 \eeq
\eq{MF1} gives a typical example of the Liouville equation which has a general
solution in the form
$N_0\Lb u;Y\Rb\,\,=\,\,N_0 \Lb Y + \phi(u) \Rb$.
 Therefore  $u$ dependence stems only from the initial conditions of  \eq{INCSD}
and for $N_0\Lb u;Y \Rb$
 we have the Balitsky-Kovchegov equation \cite{BK} in a simple form
 Namely,
\beq \label{MF2}
\frac{d N_0\Lb u;Y \Rb }{d\, \Gamma(1 \to 2)\,Y}\,\,=\,\,N_0\Lb u;Y
\Rb\,\,\,-\,\,N^2_0\Lb u;Y \Rb
\eeq
with the initial condition \cite{MUCD}
\beq \label{MF3}
N_{ic}\Lb u ,Y'_0 \Rb \,\,=\,\,N_0\Lb u;Y=Y'_0 \Rb\,\,=\,\,1\,\, -\,\,
u\,\,=\,\,\gamma
\eeq
where $\gamma$ is the amplitude at low energy.
\eq{MF2} has an obvious solution
\beq \label{MF4}
N_0\Lb u;Y \Rb\,\,\,=\,\,\,\frac{N_{ic}\Lb u ,Y'_0 \Rb\,\,\exp \Lb {\cal Y}
\Rb}{1\,\,+\,\,N_{ic}\Lb u ,Y'_0 \Rb \,\,\Lb
\exp \Lb {\cal Y} \Rb \,\,-\,\,1 \Rb}
\eeq
where ${\cal Y} \,\,=\,\,\Gamma(1 \to 2)\,Y$.  and $N_{ic}$ are the initial
conditions for the scattering amplitude
 at
$Y = Y'_0$.

To solve \eq{MF1} we can introduce  a new function   (see Ref. \cite{KL} for
details) , namely,
\beq  \label{MF5}
N_{in}\Lb u;Y,Y_0 \Rb \,\,\,\equiv\,\,\,2\,\,N_0\Lb u;Y \Rb\,\,-\,\,N_D\Lb
u;Y,Y_0 \Rb
\eeq
for which we have the same equation as \eq{MF2} but with the initial condition
given by the following equation:
\beq \label{MF6}
N^{in}_{ic}\Lb u;Y_0 \Rb \,\,\equiv\,\,N_{in}\Lb u;Y=Y_0,Y_0
\Rb\,\,=\,\,2\,N_{ic}\Lb u ,Y_0 \Rb\,\,-\,\,N^2_{ic}\Lb u ,Y_0 \Rb
\eeq

Therefore the solution to \eq{MF5} with the initial condition of \eq{MF6} has
the form (see \eq{MF4})
\bea \label{MF7}
N_{in}\Lb u;Y ,Y_0\Rb\,\,\,&=&\,\,\,\frac{N^{in}_{ic}\Lb u ,Y_0 \Rb\,\,\exp \Lb
{\cal Y} -{\cal Y}_0 \Rb}{1\,\,+
\,\,N^{in}_{ic}\Lb u ,Y_0 \Rb \,\,
\Lb \exp \Lb {\cal Y} -{\cal Y}_0  \Rb \,\,-\,\,1 \Rb} \\
 &=&\,\,\,\frac{\Lb 2\,N_{0}\Lb u ,Y_0 \Rb\,\,-\,\,N^2_{0}\Lb u ,Y_0
\Rb\,\Rb\,\,\exp \Lb {\cal Y} - {\cal Y}_0 \Rb}{1\,\,+\,\Lb 2\,N_{0}\Lb u ,Y_0
\Rb\,\,-\,\,N^2_{0}\Lb u ,Y_0 \Rb\,\Rb\,
\,\,\Lb \exp \Lb {\cal Y} -{\cal Y}_0  \Rb
  \,\,-\,\,1 \Rb} \nonumber
 \eea
and
\beq \label{MF8}
N_D\Lb u;Y,Y_0 \Rb\,\,\,=\,\,\,2\,\,N_0\Lb u;Y \Rb\,\,\,\,-\,\,\,N_{in}\Lb u;Y
,Y_0\Rb\
\eeq
Since  $  N_{in}\Lb u;Y ,Y_0\Rb$ satisfies the same equation as $ N_{0}\Lb u;Y
,Y_0\Rb$ (see \eq{MF2}) but with
a different initial condition (\eq{MF6} instead of \eq{MF3}), one can see that
the solution has the following form\cite{KL}
\beq \label{MF9}
N_{in}\Lb u;Y ,Y_0\Rb\,\,\,=\,\,\,N_0\Lb u;{\cal Y}\,\,+\,\,\Delta {\cal Y} \Rb
\,\,\,\,\mbox{with}\,\,\,\,
\Delta {\cal Y}\,\,=\,\,\ln\Lb 2 \,\,+\,\,\frac{1 -u}{u}\,\exp \Lb {\cal Y}_0
\Rb \Rb
\eeq

Using \eq{MF3} for the initial condition of $N_0\Lb u,Y=0\Rb$ and \eq{MF9}  we
obtain the following answers for the scattering amplitude,
total inelastic contribution and for the diffractive production cross section
with mass in the interval
 $ \ln (M^2/m^2)\,\,\leq\,\,Y - Y_0$:
\bea
&&N_0\Lb u;Y \Rb\,\,\,\,\,\,\,\,\,\,\,\,\,=\,\,\frac{(1 - u)\,e^{\cal Y }}{ u
\,\,+\,\,(1 - u)\,e^{\cal Y }}\,\,;
\label{S1}\\
&&N_{in}\Lb u;Y ;Y_0\Rb\,\,\,\,=\,\,\frac{\,(1 - u)\,e^{\cal Y }\,\,\Lb \,2\,u +
(1 - u)\,e^{{\cal Y }_0}\,\Rb}{ u^2 \,\,+\,\,(1 - u)\,\, e^{\cal Y }\,\,\Lb \,2
u\,\,+\,\,(1 - u)\,e^{{\cal Y }_0}\,\Rb }\,\,;\label{S2}\\
&&N_{SD}\Lb u;Y ;Y_0\Rb\,\,=\,\,\frac{\,u^2\,(1 - u)^2 \,\,e^{{\cal Y}
\,+\,{\cal Y}_0}}{
\Lb u^2\,\,+\,\,2\,u\,(1 - u)\,\,e^{{\cal Y }} \, + \,(1 - u)^2\,e^{{\cal Y}
\,+\,{\cal Y}_0}\Rb^2};\label{S3}\\
&&N_D\Lb u;Y ;Y_0\Rb\,\,\,\,=\,\,\frac{2\,(1 - u)\,e^{\cal Y }}{ u \,\,+\,\,(1 -
u)\,e^{\cal Y }}\,\,-
\,\,\frac{\,(1 - u)\,e^{\cal Y }\,\,\Lb \,2\,u + (1 - u)\,e^{{\cal Y }_0}\,\Rb}{
u^2 \,\,+\,\,(1 - u)\,\, e^{\cal Y }\,\,\Lb \,2 u\,\,+\,\,(1 - u)\,e^{{\cal Y
}_0}\,\Rb }\,\,;\label{S4}
\eea
in agreement with Refs. \cite{KL,BGLM,DGBK}.

The advantage of our way of obtaining the solution is that it  allows us to take
into account an arbitrary initial conditions.
We will use this in the more general case than
the mean field approach.

\subsection{Diffractive production beyond the mean field approximation}
In the kinematic region of rather small produced mass (see \eq{MR}) but at very
high energies (see \eq{MR1})
 we can neglect the correlations
  \cite{IM,KOLE,SHBOW} in the diffractive produced system due to the Pomeron
loops (see \fig{cutps}-c, for example).

\DOUBLEFIGURE[h]{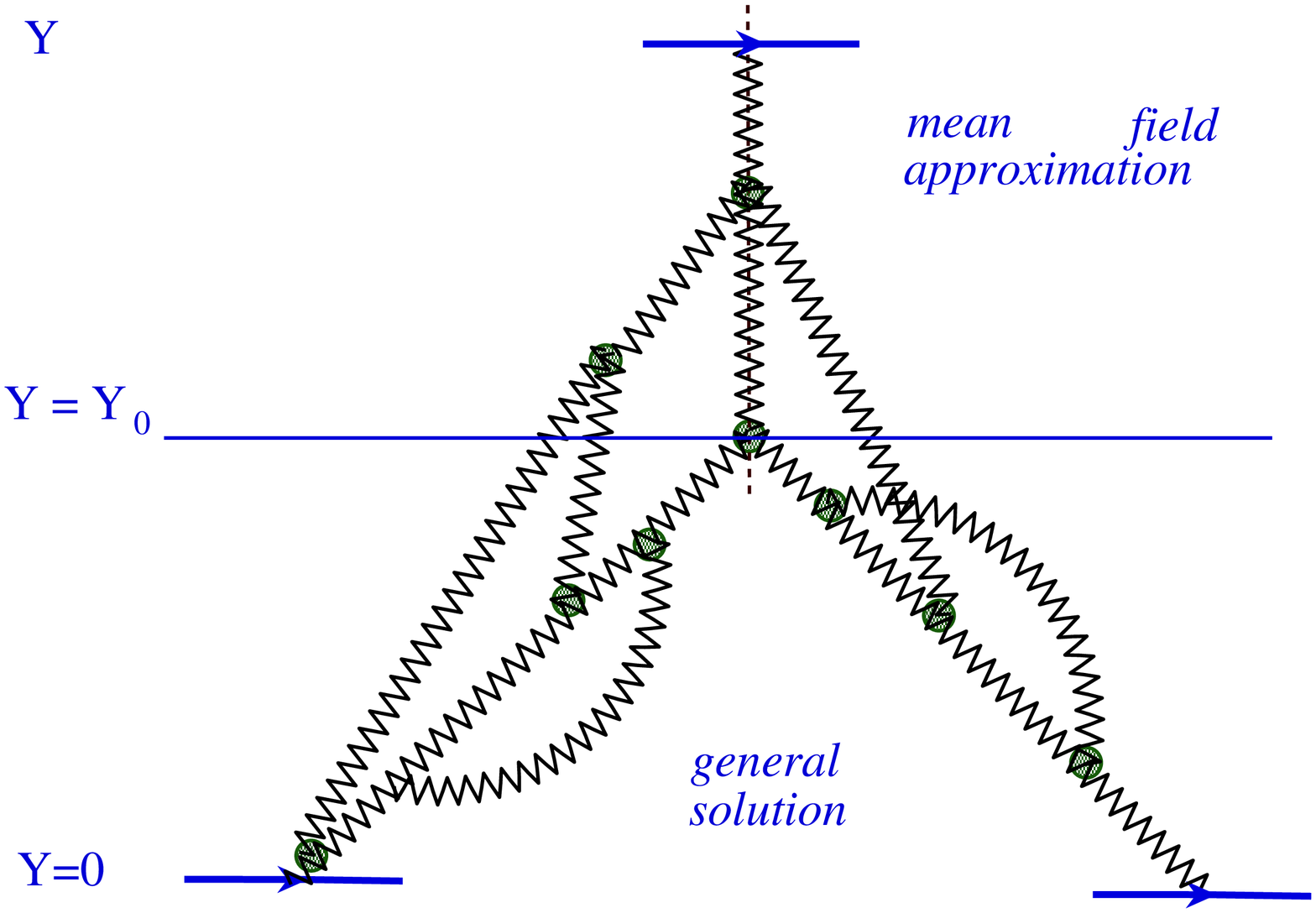,width=70mm,height=60mm}{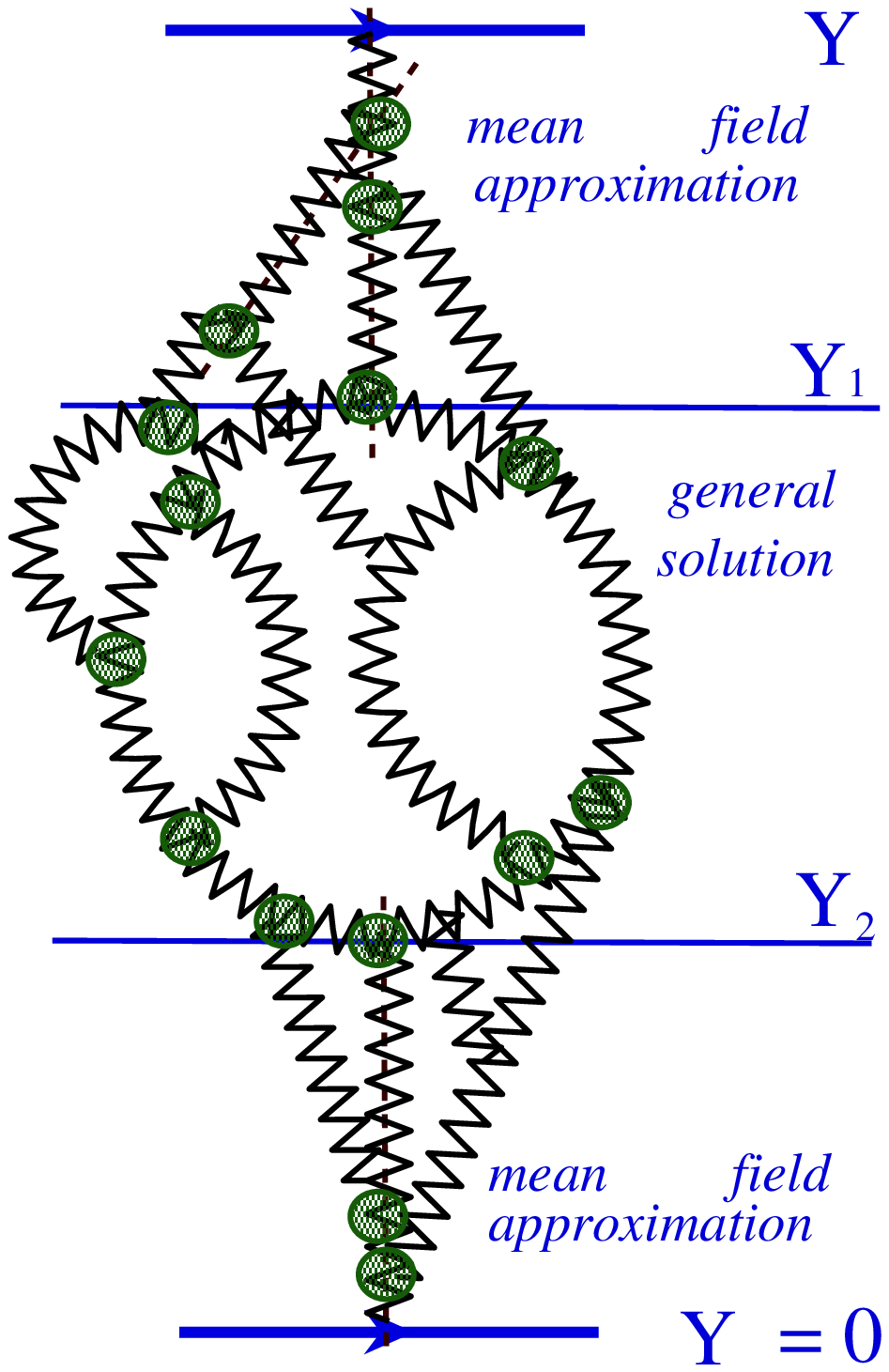,width=45mm,height=55mm}
{The general diagram for single diffraction production in the kinematic region
given by \protect\eq{MR} and
\protect\eq{MR1}
\label{sdgen}. The zigzag line crossed by the dotted line denotes the cut
Pomeron
  while the same line without the dotted one stands for the Pomeron
exchange.}{The general diagram for double
diffraction production
in the kinematic region, given by \protect\eq{MR}, for both masses
$\ln (M^2_1/m^2)\,\,=\,\,Y- Y_1$ and $\ln (M^2_2/m^2)\,\,=\,\,Y_2$
 and in the energy region , given by  \protect\eq{MR2}. The zigzag line crossed
by the dotted line denotes the cut Pomeron
  while the same line without the dotted one stands for the Pomeron exchange.
\label{ddgen}
}

  The only diagrams that we need to consider are shown in \fig{sdgen}.  Our
solution, given by \eq{MF7},  sums
these diagrams
if we substitute in \eq{MF7} the exact solution
   of the general equation for $N_{in}\Lb u;Y_0\Rb =\,\,2\,\,N_0\Lb u;Y_0\Rb$
that includes the Pomeron loops.
 This equation has the form \cite{L3,KLTM}
\beq\label{GA1}
\frac{\partial N_{0}\Lb u;{\cal Y}_0 \Rb}{\partial{\cal Y}_0}\,\,\,\,= \,\,\,\,
u\,(1 - u)\,\left(-\,\,\,\,
\frac{\partial
N_{0}\Lb u;{\cal Y}_0\Rb}{\partial
u}\,\,+\,\,\frac{1}{\kappa}
\frac{\partial^2 N_{0}\Lb u;{\cal Y}_0\Rb}{\partial u^2} \right)
\eeq
where we denote ${\cal Y}\,\,=\,\,\Gamma(1 \to 2)\,\,Y$ and
$\kappa\,\,=\,\,\Gamma(2\to 1)/\Gamma(1 \to 2)$.
$\Gamma(1 \to 2)$ and $\Gamma(2 \to 1)$ describe the $ P \to 2 P$ and $2 P \to
P$  transition, respectively.

The initial  and boundary conditions are
\bea
&&\mbox{Initial conditions:}\,\,N_{0}\Lb u;{\cal Y}_0=0 \Rb\,\,=\,\,1 - u\,\,;
\label{IC}\\
&&\mbox{Boundary conditions:}\,\,N_{0}\Lb u = 0 ;{\cal
Y}_0\Rb\,\,=\,\,1\,\,;\,\,\,\,\,\,\,N_{0}\Lb u = 1;{\cal Y}_0\Rb\,\,
=\,\,0 \,\,; \label{BC}
\eea
\eq{IC} stems from the fact that at low energies only  the  exchange of one
Pomeron contributes  to the scattering amplitude.
The boundary condition at
$u = 1$ says that the dipole at high energies does not interact with the target
if this dipole interacts with
 zero amplitude at low energies ( $Y_0 =0$).  At $u =0$ the dipole interacts at
low energy with the amplitude which is equal
to the maximal value
 ($N = 1$) and any interaction between Pomerons cannot change this value of the
amplitude.

\FIGURE{
\centerline{\epsfig{file=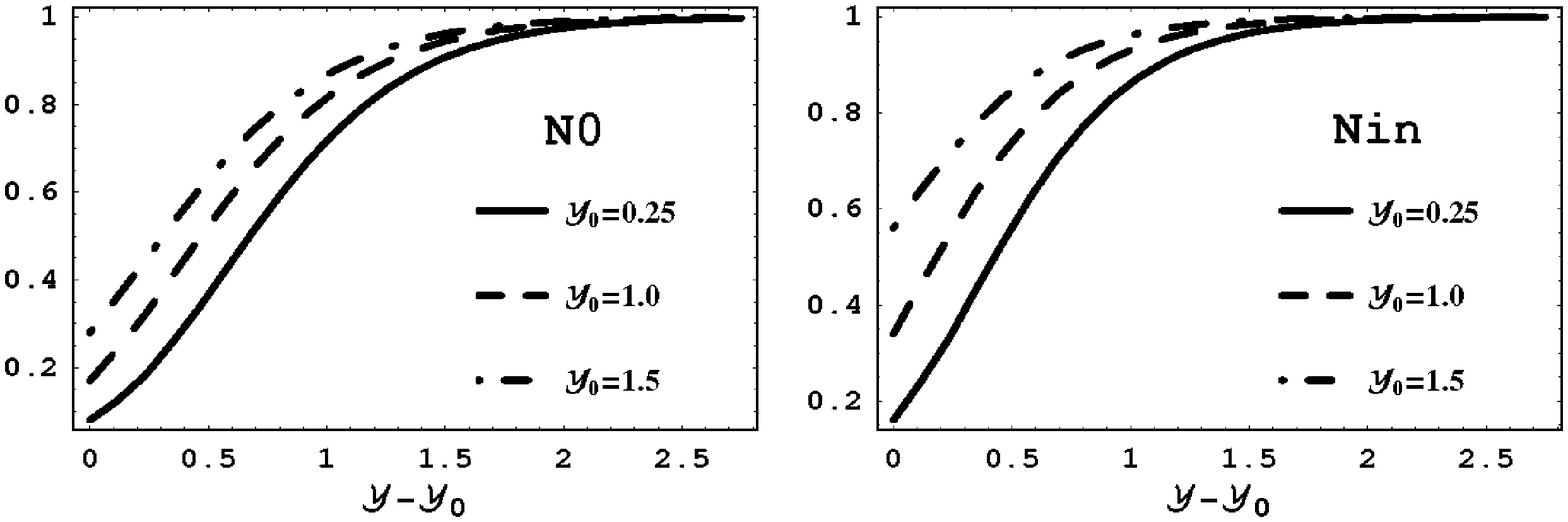,width=170mm}}
\caption{The amplitudes $N_0\Lb 1 - u=1/\kappa, Y; Y_0 \Rb$  and $N_{in}\Lb 1 -
u=1/\kappa, Y; Y_0 \Rb$
as function of ${\cal Y} - {\cal Y}_0 \,=\,\,\Gamma(1 \to 2)\cdot(Y - Y_0)$ at
different values of ${\cal Y}_0$}
\label{n0in}
}

\FIGURE{
\centerline{\epsfig{file=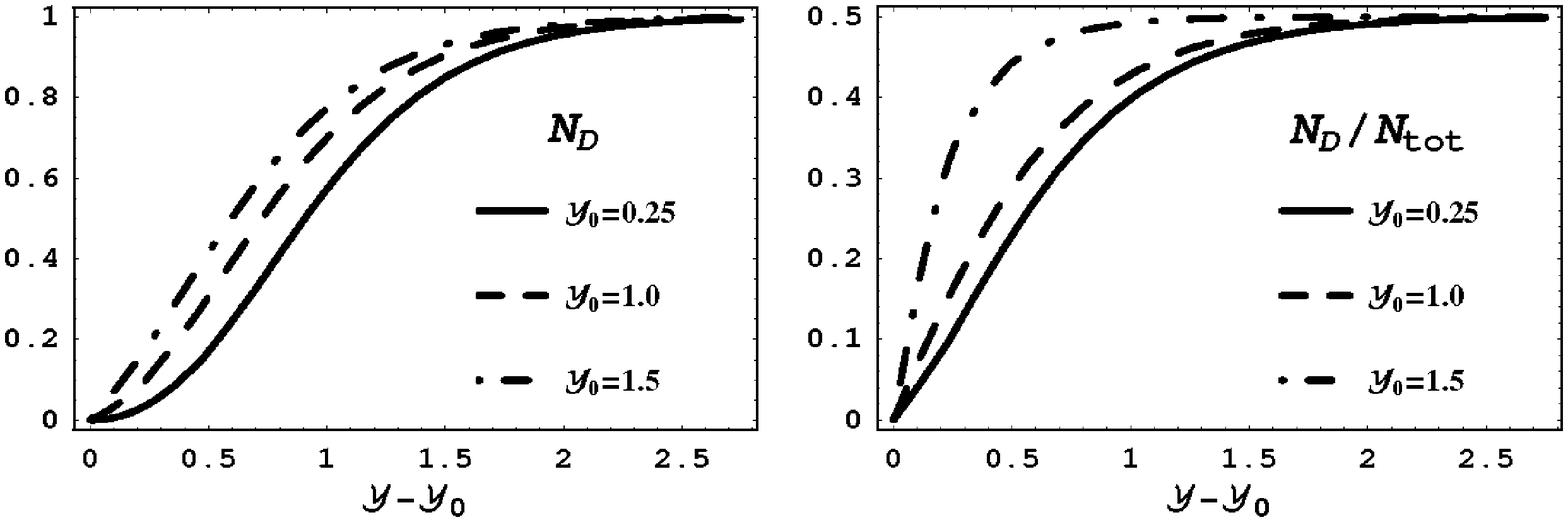,width=170mm}}
\caption{The total cross section of the diffraction dissociation in the range of
mass (M) $ ln (M^2 /m^2)\,\,\leq\,\,Y_0$ $N_D\Lb 1 - u=1/\kappa, Y; Y_0 \Rb$ 
and the ratio of this cross section to the elastic amplitude ( the total cross
section is equal to 2 $N_0\Lb 1 - u=1/\kappa, Y \Rb$)
as function of ${\cal Y} - {\cal Y}_0 \,=\,\,\Gamma(1 \to 2)\cdot(Y - Y_0)$ at
different values of ${\cal Y}_0$}
\label{nd}
}

\FIGURE{
\centerline{\epsfig{file=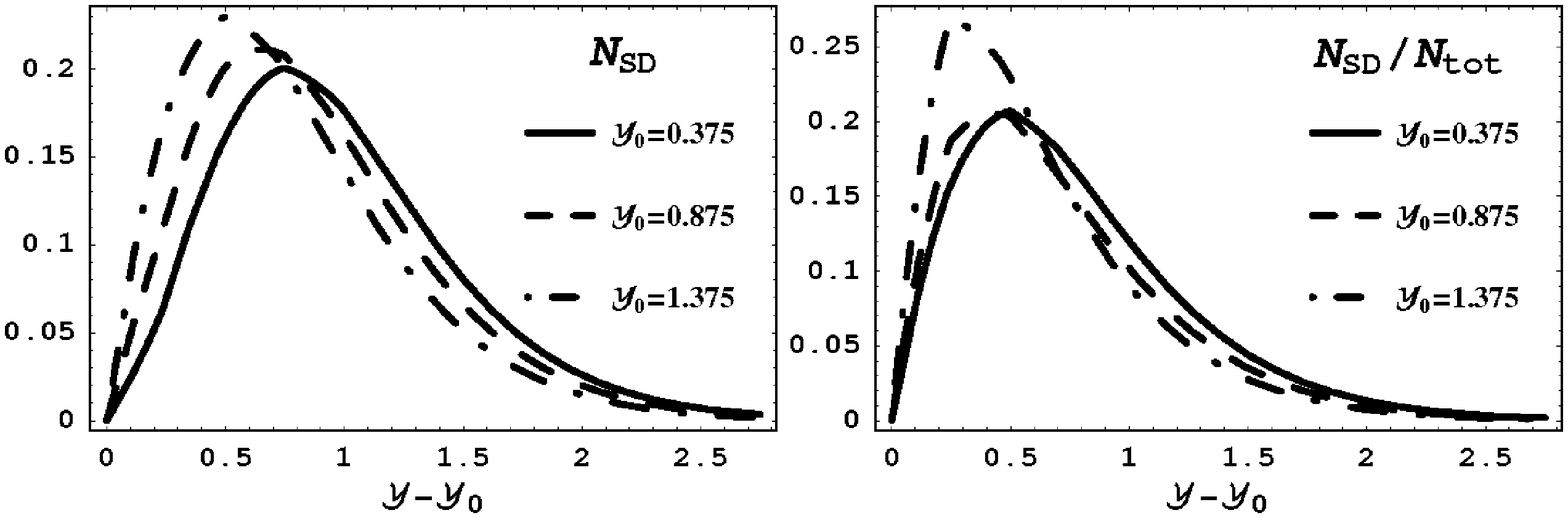,width=170mm}}
\caption{The  cross section of the diffraction dissociation at in the range of
mass ${\cal Y}_{0,i} +\frac{ \Delta {\cal Y}_0 }{2} \,-\, \Lb  {\cal Y}_{0,i}
-\frac{ \Delta{\cal Y}_0 }{2} \Rb = \Delta{\cal Y}_0 $ determined by 
$N_{SD}\Lb 1-u= 1/\kappa;Y ;Y_{0,i}\Rb\,=\,\,\Lb N_{D}\Lb 1-u= 1/\kappa;Y
;Y_{0,i} +\frac{ \Delta{\cal Y}_0 }{2} \Rb \,-\,N_{D}\Lb 1-u= 1/\kappa;Y
;Y_{0,i} -\frac{ \Delta{\cal Y}_0 }{2} \Rb \Rb/\Delta {\cal Y}_0$
with $\Delta {\cal Y}_0 = 0.25$  and the ratio $N_{SD}\Lb 1-u= 1/\kappa;Y
;Y_{0,i}\Rb/N_0\Lb 1 - u=1/\kappa, Y \Rb$
at different values of ${\cal Y}_{0,i}$.}
\label{nsd}
}
The solution for the single diffractive production can be found within the
following procedure:

\begin{enumerate}
 \item \quad For the cross section of diffractive production of the mass $M$
($Y = \ln M$) the  initial conditions
 for the solution of the master equation (see \eq{GA1}) we put at $Y =Y_0$ and
they have a form
 \bea
 N_0\Lb u,Y_0 \Rb \,\,&=& \,\,\,\frac{(1 -u)\,e^{{\cal Y}_0}}{1\,\,+\,\,(1 -
u)\,(e^{{\cal Y}_0} \,-\,1)}; \label{SDIC1}\\
 N_{in}\Lb u,Y_0 \Rb \,\,&=& \,\,\,\frac{2\,(1 -u)\,e^{{\cal
Y}_0}}{1\,\,+\,\,\,(1 - u)\,(e^{{\cal Y}_0} \,-\,1)}\,\,
 -\,\,\Lb\frac{(1 -u)\,e^{{\cal Y}_0}}{1\,\,+\,\,(1 - u)\,(e^{{\cal Y}_0}
\,-\,1)}\Rb^2;
 \label{SDIC2}
 \eea
\item \quad The next step is to find the exact solution ($ N^{exact}_0\Lb
u,Y;Y_0 \Rb $ and $ N^{exact}_{in}\Lb u,Y;Y_0 \Rb $ )
to  \eq{GA1} with the initial conditions determined by \eq{SDIC1} and
\eq{SDIC2}. The general method for finding such solution is given in Ref.
\cite{KLTM} and the only change that we need to make here is to include
different initial condition.
\item \quad
However, in  Ref. \cite{KLTM} we use Kovchegov's formula \cite{BK,L2} for the
amplitude, namely,
\beq \label{KAMP}
N_0\Lb 1 - u;Y \Rb\,\,\,=\,\,1 \,\,-\,\,Z\Lb  u;Y \Rb\
\eeq
where $Z$ is the generating function. This formula assumes that $n$-dipoles can
interact with one dipole (target) independently.
 It sums all diagrams of \fig{cutps}-b type. However\footnote{We thank Edmond
Iancu and Al Mueller for hot discussions on this subject.},  if
we would like to restrict ourselves by the enhanced
  diagrams of \fig{sdgen} we need to use a different relation between $Z$ and
$N$.  Namely,
\bea
N_0\Lb 1 - u=1/\kappa;Y;Y_0 \Rb\,\,&=&\,\,\,\frac{\partial\,Z\Lb  u;Y
\Rb}{\partial\,u}|_{u=1}\,\frac{1}{\kappa} \,;\label{N0}\\
N_{in}\Lb 1 - u=1/\kappa;Y;Y_0 \Rb\,\,&=&\,\,\,\frac{\partial\,Z_{in}\Lb  u;Y
\Rb}{\partial\,u}|_{u=1}\, \frac{1}{\kappa}\,;\label{NIN}
\eea
In \fig{n0in} the solution given by \eq{N0} and \eq{NIN} is plotted as a
function of $Y - Y_0$ at different values of $Y_0$.
\item \quad
 Using \eq{N0} and \eq{NIN} we can
obtain the solution for the single diffraction based on \eq{MF8}, namely,
\beq
N_D\Lb 1- u = 1/\kappa;Y ;Y_0\Rb\,\,\,\,=\,\,2\,N_0\Lb 1 - u=1/\kappa;Y;Y_0
\Rb\,\,-\,\,N_{in}\Lb 1 - u = 1/\kappa;Y;Y_0 \Rb
\eeq
and for the cross section of single diffraction in the system with mass $Y_0 =
\ln M$ we have
\beq \label{NSD}
N_{SD}\Lb 1-u= 1/\kappa;Y ;Y_0\Rb\,\,=\,\,-\,\frac{d\,N_D\Lb 1- u = 1/\kappa ;Y
;Y_0\Rb}{d Y_0}
\eeq
\end{enumerate}
Our calculations for $N_D\Lb 1- u = 1/\kappa;Y ;Y_0\Rb$ and $N_{SD}\Lb 1-u=
1/\kappa;Y ;Y_0\Rb$
are shown in \fig{nd} and \fig{nsd}. The most interesting observation is the
fact that in a wide range of produced mass ${\cal Y}_0 <1$ these cross sections
only weakly depend on the value of the mass.

\section{Double diffractive production}
\label{sec:DD}
In this section we will discuss the double diffractive production, or the
reaction of the following type
\bea \label{DDR}
&&\mbox{\{hadron (dipole)\}\,\,\, +\,\,\,\{ hadron (dipole)\}}
\,\,\,\Longrightarrow \,\,\, \\
&& \mbox{\{ hadron(gluon) system with mass $M_1$\}}\,\,\,\,+\,\,\,\,
\mbox{\{LRG\}}\,\,\,\,+\,\,\,\,\mbox{\{ hadron(gluon) system with mass $M_2$}\}
\nonumber
\eea
where both masses ($M_1$ and $M_2$) satisfy the relation of \eq{MR1} while the
total energy is so large that \eq{MR2}
 holds.
The diagrams that contribute to this process in the kinematic region, that we
have mentioned, are shown in \fig{ddgen}.
For the double diffraction production in the kinematic region, given by \eq{MR},
for both masses
$\ln (M^2_1/m^2)\,\,=\,\,Y- Y_1$ and $\ln (M^2_2/m^2)\,\,=\,\,Y_2 - 0$
 and in the energy region, given by  \eq{MR2}, we can neglect the Pomeron loops
contributions in the region
 $Y - Y_1$ and $Y_2 -0$
 (see \fig{ddgen}).   This simplification leads to a simple procedure of
calculating the cross section for the
double diffractive dissociation,
 which consists of  four  steps:
 \begin{enumerate}
 \item \quad As the initial conditions for the double diffractive processes we
choose
\eq{SDIC1} and \eq{SDIC2} at $Y_0 = Y_2$. They give us $N_{in}\Lb u,Y_2\Rb$ and
$N_{0}\Lb u,Y_2\Rb$;

\item \quad Using the general method of Ref. \cite{KLTM},  we find the
generating function $Z\Lb u, Y_1 - Y_2;Y_2\Rb$
and $Z_{in}\Lb u, Y_1 - Y_2;Y_2\Rb$
as   the solution of \eq{GA1} with the initial conditions
\bea
Z_{in}\Lb u, Y_1 - Y_2=0;Y_2\Rb\,\,\, &=&\,\,\,1\, \,-\,\,N_{in}\Lb u,Y_2\Rb\,;
\label{DDIC1}\\
Z\Lb u, Y_1 - Y_2=0;Y_2\Rb\,\,\,&=&\,\,\,1 \,\,-\,\,N_{0}\Lb u,Y_2\Rb\,;
\label{DDIC2}
\eea
\item \quad The third step is to notice that the amplitudes $N_{0}\Lb u,Y -
Y_1\Rb$ and  $N_{in}\Lb u,Y - Y_1\Rb$
are also determined by \eq{SDIC1} and \eq{SDIC2} at $Y_0 = Y - Y_1$

\item \quad The fourth
 and the last step is to find the final answer using the general formula (see
Ref.\cite{IM,KOLE,L1,L2} ) that was re-written in the compact form  in Ref.
\cite{KODD}
\bea
&&N_0\Lb \gamma=1/\kappa,Y-Y_1;Y_1;Y_2\Rb\,\,\,\,=\label{DDN0}\\
&&1\,\,\,-\,\,\left\{\exp \Lb -\,\gamma\,\frac{d}{d \,u_{1}}\,
\frac{d}{d \,u_{2 }}\Rb\,\,\,
N_0\Lb Y-Y_1,u_1 \Rb\,\,Z\Lb Y_1;Y_2\,,u_2 \Rb \,\right\}\mid_{u_{1 } =1; \,u_{2
}=1}\,; \nonumber\\
&&N_{in}\Lb \gamma = 1/\kappa,Y-Y_1;Y_1;Y_2\Rb\,\,\,\,=\label{DDNIN}\\
&&1\,\,\,-\,\,\left\{\exp \Lb -\,\gamma\,\frac{d}{d \,u_{1 }}\,
\frac{d}{d \,u_{2 }}\Rb\,\,\,
N_{in}\Lb Y-Y_1,u_1 \Rb\,\,Z_{in}\Lb Y_1;Y_2\,,u_2 \Rb \,\right\}\mid_{u_{1 }
=1; \,u_{2 }=1}\,;\nonumber
\eea
\end{enumerate}
We can simplify \eq{DDN0} and \eq{DDNIN} going to the Mellin transform for the
generating functions $Z_{in}\Lb Y_1;Y_2\,,u_2 \Rb $ and $Z\Lb Y_1;Y_2\,,u_2 \Rb
$. Namely,
\beq \label{MTRA}
Z\Lb Y_1;Y_2\,,u \Rb\,\,\,=\,\,\,\int^{a + i \infty}_{a - i
\infty}\,\,\frac{d\,\mu}{2 \pi\,i}\,e^{\mu\, \ln(1/u)}\,\tilde{Z}\Lb
Y_1;Y_2\,,\mu \Rb
\eeq
Since $d^n N_0\Lb Y-Y_1,u_1 \Rb/d^2 u_1$ is known, namely,
\beq \label{DERV}
\frac{d^n N_0\Lb Y-Y_1,u_1 \Rb}{d u^n_1}\,\,\, =\,\,n!\,\,
e^{{\cal Y} - {\cal Y}_1}\,\times\,\Lb e^{{\cal Y} - {\cal
Y}_1}\,\,-\,\,1\,\Rb^{n -1}
\eeq
 \eq{DDN0} reads as
\bea \label{AMPMU}
&&N_0\Lb \gamma,Y-Y_1;Y_1;Y_2\Rb\,\,= \\
&&\int^{a + i \infty}_{a - i \infty}\,\,\frac{d\,\mu}{2 \pi i}
\,\,\tilde{Z}\Lb Y_1;Y_2\,,\mu \Rb\,\,\frac{e^{{\cal Y} - {\cal Y}_1}}{\Lb
e^{{\cal Y} - {\cal Y}_1}\,\,-\,\,1\,\Rb}
\,\,\left\{-\,\Gamma\Lb 1 - \mu,\frac{1}{\gamma T} \Rb\,\Lb
\frac{1}{\gamma\,T}\Rb^{\mu}\,\exp\Lb \frac{1}{\gamma\,T}\Rb \,\,+\,\,1
\right\} \nonumber
\eea
where $T\,\equiv\,\Lb e^{{\cal Y} - {\cal Y}_1}\,\,-\,\,1\,\Rb$ and $\gamma =
1/\kappa$.
Since (see Ref. \cite{RY} formula {\bf 3.381(3)})
\beq \label{GAIN}
\int^{\infty}_1\,\,d\,x \,x^{- \mu}\,\exp \Lb -
\frac{x}{\gamma\,T}\Rb\,\,=\,\,\Lb \frac{1}{\gamma\,T}\Rb^{\mu -1}\,\Gamma\Lb 1
- \mu,\frac{1}{\gamma\,T} \Rb
\eeq
one can see that \eq{AMPMU} can be re-written in the form
\bea \label{AMPMU1}
&&N_0\Lb \gamma,Y-Y_1;Y_1;Y_2\Rb\,\,\times\,\,\Lb \frac{e^{{\cal Y} - {\cal
Y}_1}\,\,-\,\,1\,}{e^{{\cal Y} - {\cal Y}_1}}\Rb\,\,\,= \\
&&-\,\,\gamma\, T\,\exp\Lb \frac{1}{\gamma\,T}\Rb\,\,\int^{\infty}_1\,\,d\,x\,\,
\exp \Lb - \frac{x}{\gamma\,T}\Rb \,\int^{a + i \infty}_{a - i
\infty}\,\,\frac{d\,\mu}{2 \pi i}\,\,x^{- \mu}\,\,\tilde{Z}\Lb Y_1;Y_2\,,\mu
\Rb  \,\, +\,\,1\,\, =\nonumber\\
&&-\,\,\gamma T\,\exp\Lb
\frac{1}{\gamma\,T}\Rb\,\,\int^{\infty}_1\,x\,d\,x\,\exp \Lb -
\frac{x}{\gamma\,T}\Rb\,\,
Z\Lb Y_1;Y_2\,,\ln(1/x) \Rb\,\,+\,\,1\,=\nonumber\\
&&-\,\,\gamma T\,\exp\Lb
\frac{1}{\gamma\,T}\Rb\,\,\int^1_0\,\frac{d\,u}{u^2}\,\exp \Lb -
\frac{1}{u\,\gamma\,T}\Rb\,\,Z\Lb Y_1;Y_2\,,u \Rb\,\,+\,\,1 \nonumber
\eea
\eq{AMPMU1} is convenient for calculation. In derivation of this equation we
used that $Z\Lb Y_1;Y_2\,,u=1\Rb\,\,=\,\,1$ (see \eq{BC}).

For $N_{in}\Lb \gamma,Y-Y_1;Y_1;Y_2\Rb$ we obtain a reasonable approximation by
using the same expression as \eq{AMPMU1} but with
$ T \,\to 2\,T$ and $Z \,\to Z_{in}$. Indeed,  in \eq{SDIC2} we can consider $1
- u\,\approx\,1/\kappa\,\,\,\ll\,\,1$
and neglect the second term in  \eq{SDIC2} if  $( 1 - u)\,\exp( {\cal Y} - {\cal
Y}_0)\,\,\ll\,\,1$ or
$ {\cal Y} - {\cal Y}_0\,\ll\,1/(1 -u) \approx \kappa$. Recall that this
interval for $ {\cal Y}  - {\cal Y}_0$  is much wider than the one given by
\eq{MR}, since $\kappa = 1/\as^2$.

Finally,
the inclusive double diffractive process, namely, the cross section defined as
\beq \label{DEFINDD}
N^{DD}_{in}\,\,=\,\,\int^{Y - Y_1}_0\,\,d
\,\ln(M^2_1/m^2)\,\,\int^{Y_2}_0\,\,d\,\ln(M^2_2/m^2)\,\,
M^2_1\,M^2_2\,\frac{d \sigma}{d\, M^2_1\,d\,M^2_2}
\eeq
we calculate as
\beq \label{STP4}
N^{DD}_{in}\Lb u;Y ,Y_1,Y_2\Rb\,\,\,=\,\,2\,N_0 \Lb u,Y\Rb
\,\,\,-\,\,\,N_{in}\Lb u;Y ,Y_1,Y_2\Rb
\eeq
  For calculation of the double diffraction cross section in the two system of
hadrons with fixed mass
we need
\beq \label{DDXS}
M^2_1\,M^2_2\,\frac{d \sigma}{d\, M^2_1\,d\,M^2_2}\,\,=\,\,-\,\frac{
d\,N^{DD}_{in}\Lb u;Y,Y_1,Y_2\Rb}{d\,Y_1\,d\,Y_2}
\eeq

\section{Survival probability for  the   central diffractive production.}

In our approach the value, that we call survival probability, is defined as the
following ratio:
\beq \label{SP1}
<|S^2|>\,\,\,\,=\,\,\,\frac{| A\Lb \fig{hp}-a\Rb|^2}{|A\Lb \fig{hp}-b\Rb|^2}
\eeq
This ratio is the quantitative measure how the interactions of the `wee' partons
from   different parton showers are suppressed the probability of the production 
of the central system, which is produced with rapidities close to $Y/2$ in the 
reaction of \eq{MR3}  type.

\DOUBLEFIGURE[ht]{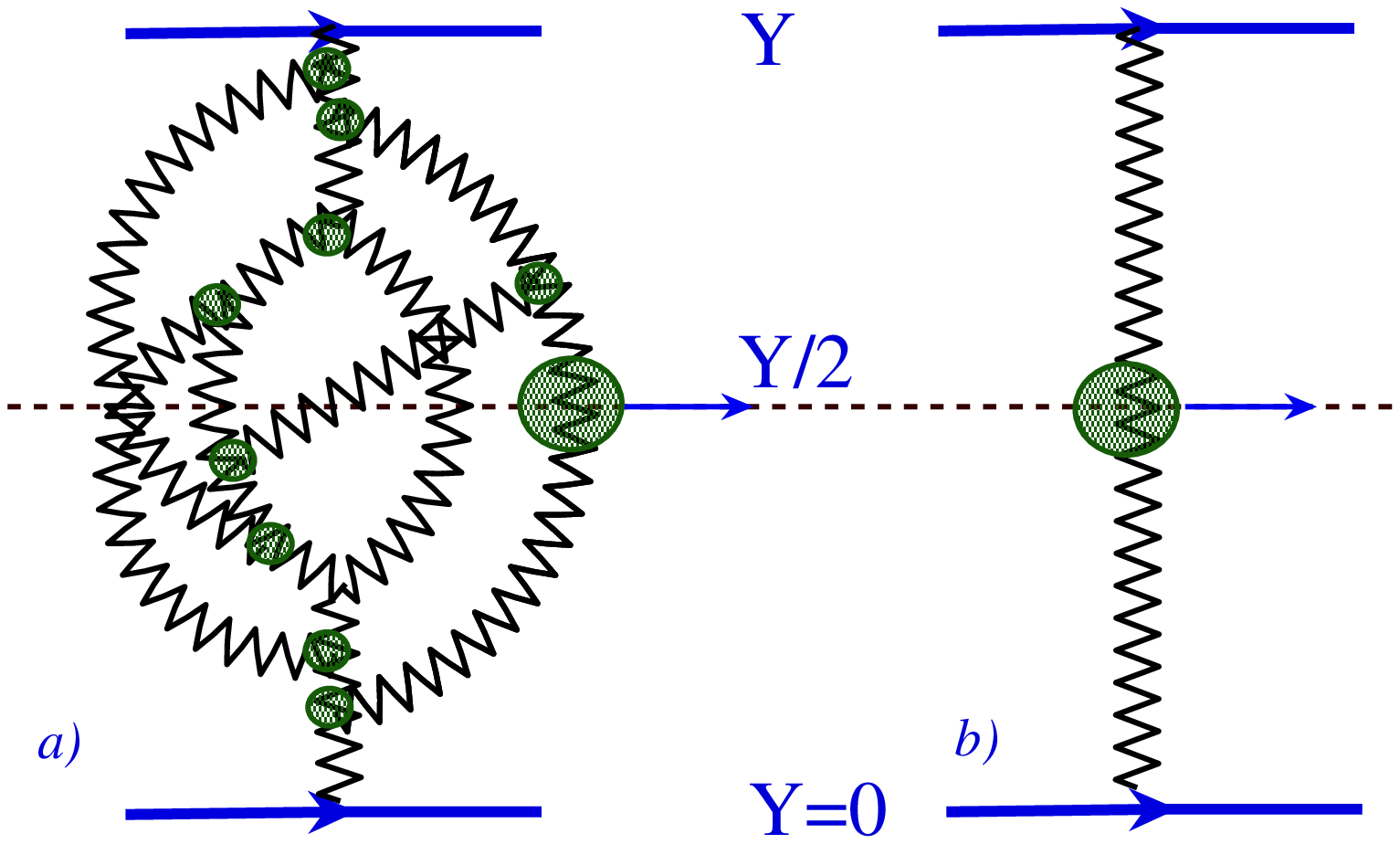,width=80mm,height=50mm}{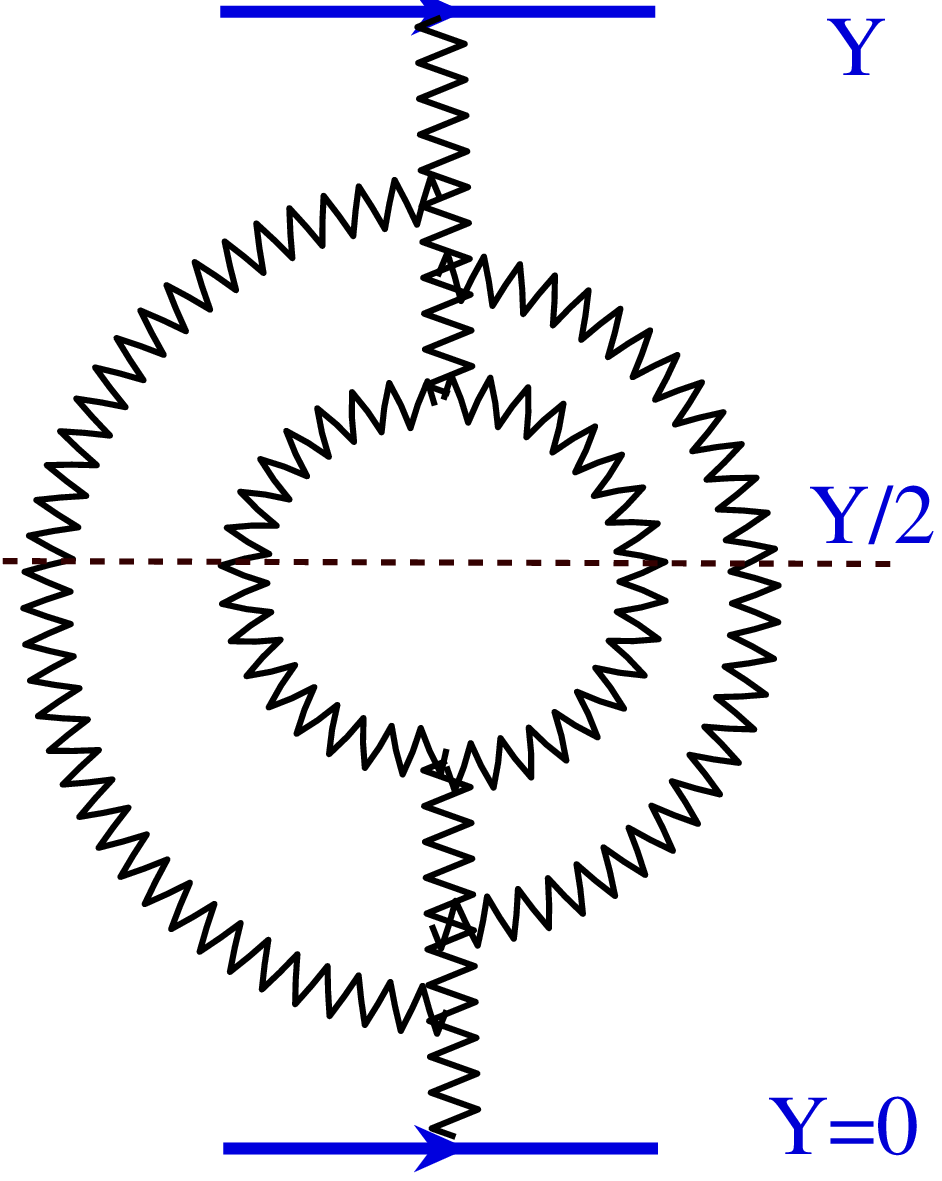,width=50mm,height=47mm}
{The central diffractive Higgs meson production: diagrams \protect\fig{hp}-a is
a general case that takes into account
the interaction of many parton showers, while the  diagram of \protect\fig{hp}-b
is the central production from one
parton shower. \label{hp}}{ The diagrams that contribute in
Mueller-Patel-Salam-Iancu approximation, which describe
the amplitude for the rapidities that satisfy  the restriction
 $Y\,\,\equiv\,\,\ln\left( M^2/m^2 \right)\,\,\ll \,\,\frac{1}{\bas}\,
\ln \left( \frac{N^2_c}{\bas^2} \right)$ . \label{mpsi}}

To find the amplitude for the diagram of \fig{hp}-a we need to take into
account that the Higgs meson in \fig{hp} can be emitted from any of Pomerons  
at rapidity $Y/2$ in this diagram.  We can take into account this fact by the 
following procedure.
 First, we recall that the scattering amplitude can be written in the form
\cite{L1,L2,BK}
 \beq \label{SP2}
 N\Lb \gamma; \frac{Y}{2} \Rb\,\,\,=\,\,-
\sum^\infty_{n=1}\,\,\frac{(-1)^n}{n!}\,\gamma^n \,\frac{\partial^n Z\Lb u,Y\Rb
}{\partial^n u}|_{u =1} \,\,
 =\,\,- \sum^\infty_{n=1}\,\,(-1)^n\,\gamma^n\,\,\,\rho_n\,\,
\eeq
Therefore, to calculate the Higgs production we need to take the following
derivative
\beq \label{SP3}
\gamma \frac{\partial  N\Lb \gamma; Y \Rb}{\partial \gamma}\,\,=\,\,-
\sum^\infty_{n=1}\,\,n\,(-1)^n\,\gamma^n\,
\,\,\rho_n\,\,
\eeq
The next step is to solve our master equation (see \eq{GA1})  using \eq{SP3} as
the initial condition. This solution
($N\Lb \gamma; Y; \frac{Y}{2} \Rb$)
 will allow us to  find the survival probability.
For calculating the survival probability we need to use the equation
\beq \label{SP4}
<|S^2|>\,\,\,\,=\,\,\,\Lb \frac{N\Lb \gamma; Y; \frac{Y}{2} \Rb}{\gamma \,\exp
\Lb {\cal Y} \Rb} \Rb^2
\eeq

\FIGURE{
\centerline{\epsfig{file=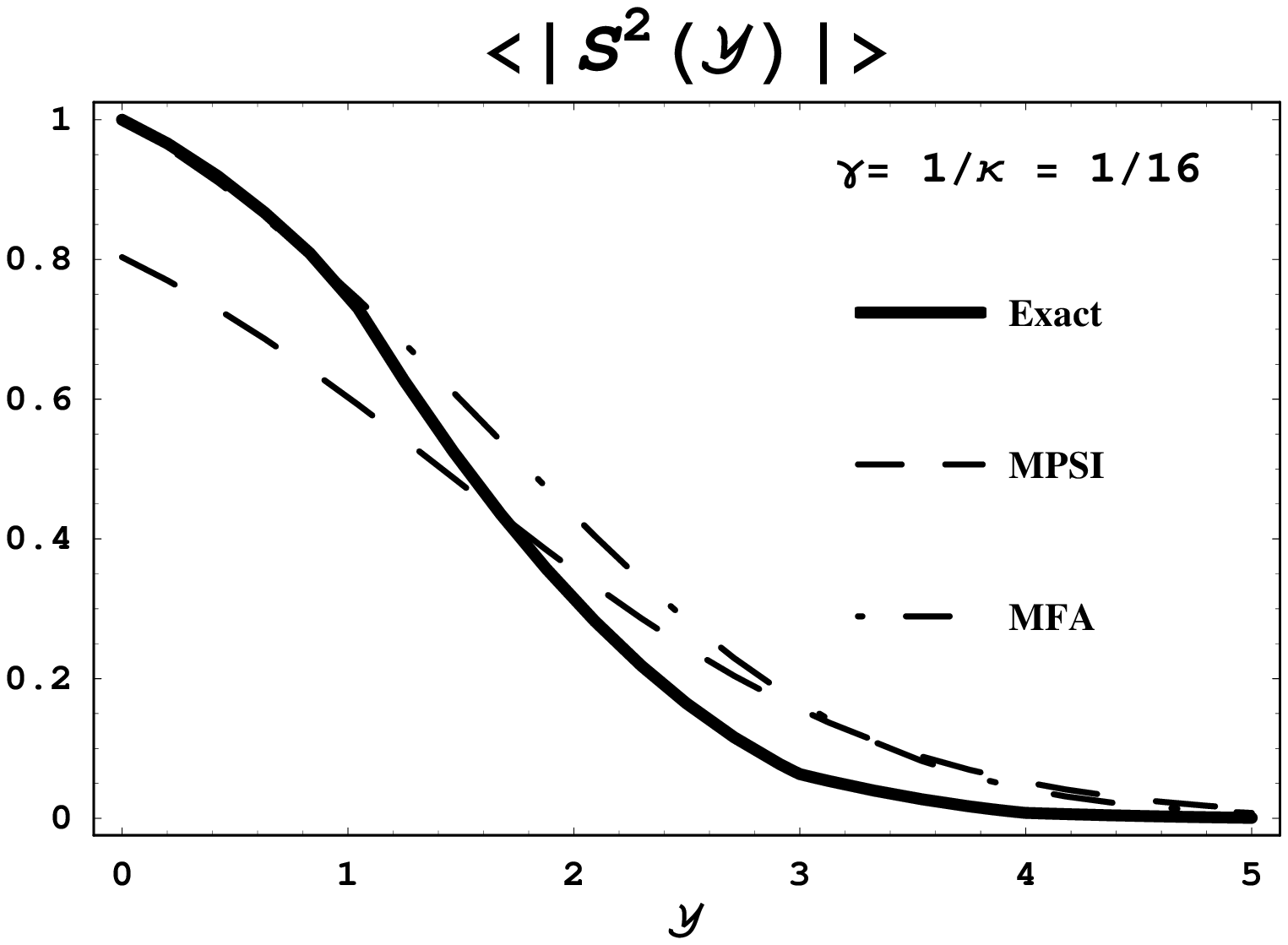,width=140mm}}
\caption{ Survival probability $\langle|S^2\rangle$  versus ${\cal Y} \equiv
\Gamma(1\to 2)\,Y$ at different values for the low energy amplitude $\gamma$.
The solid line is our exact calculation (\eq{SP4}) while the dotted line is the MPSI
approximation (\eq{CPS6}) and the dashed-dotted line shows the calculation in the MFA (\eq{CP3})}
\label{sp}
}

In \fig{sp} our estimates for the value of the survival probability is plotted
versus  energy.
We need to know the value of the $ P \to 2P$ vertex to assign the measured value
of the total rapidity $Y = \ln(s/s_0)$ with $s_0 \approx 1 \,GeV^2$. Since this
value is very important characteristic of the diffractive Higgs  production we
present the result of calculation in Table 1. If we take $\Gamma(1 \to 2) =
\Delta_P(0)$ where $\Delta_P(0)$ is the intercept of soft Pomeron 
($\Delta_P(0)\,\approx \,0.1$) the value of $\langle|S^2|\rangle \approx $ 0.23.
 This value is in agreement with the estimates based on the two channel model for
diffractive production \cite{DG,GLMSB}.
 However, if $\Delta_P(0)\,\approx \,\bas = 0.25$  the value of
$\langle|S^2|\rangle \, =\, 7\,\cdot\, 10^{-4}$ which is much smaller.
It should be stressed that our calculations are one of the first that includes
the enhanced diagrams (Pomeron loops) and can be used to evaluated the effect
of this contribution on the value of the survival probability. One can see that
the effect can be very large and the process of the diffractive Higgs
production could be much less than it is expected
based on the estimates in Refs. \cite{DG,DGK,GLMSB}.

\TABLE[ht]{
\begin{tabular}{|l|l|l|l|l|l|l|l|l|l|l|}
  \hline\hline
 ${\cal Y}$ & 0 &  0.25 & 0.5 & 0.75 & 1. & 1.25
  & 1.5 & 1.75 & 2. & 2.25 \\
  \hline
  $\langle|S^2|\rangle$ & 1&  0.96 & 0.83 & 0.75 &  0.62 &  0.751&  0.41  &
0.31 & 0.23 &  0.16\\
\hline \hline
 ${\cal Y}$ & 2.5 &   2.75 & 3. & 3.25 & 3.75 & 4.
  & 4.25 & 4.5 &4.75. & 5.  \\
  \hline
  $\langle|S^2|\rangle$  & 0.11 & 0.063 &  0.045 &  0.029 & 0.017 & 0.0077 &  0.0054 &  0.0034 &  0.0018 &  0.0007  \\
\hline \hline
  \end{tabular}
  \caption{The values of the survival probabilities versus ${\cal Y}\,=\,\Gamma(1 \to 2)\,Y$ at the value of  the low energy amplitude $\gamma = 1/\kappa$.}}

The estimates, discussed above,  have one obvious defficientcy: they do not take into account the fact
that  the Higgs meson production cannot occur for energies less than $s = M^2$ where $M$ is the mass if
Higgs meson.
In other language , the vertex for Higgs meson production has a size in rapidity while we
neglected this effect in \eq{SP4}.  It is easy to take into account this replacing \eq{SP4} by
\beq \label{SP44}
<|S^2|>\,\,\,\,=\,\,\,\Lb \frac{N\Lb \gamma; Y; \frac{Y}{2} \Rb}{ N\Lb \gamma; \zeta; \frac{\zeta}{2} \Rb
}\Rb^2\,\,\,\,\mbox{with}\,\,\,\zeta\,\,=\,\,\ln \Lb \frac{M^2_{Higgs}}{s_0} \Rb
\eeq

In the Table 2 we show values of $<|S^2|>$ for $\bas = 0.25$, when calculations ware archived using \eq{SP44}. One can see that
the difference between \eq{SP4} and \eq{SP44} (Table 1 and Table 2 correspondingly) is  large and expected $ <|S^2|>$ at the LHC energy
is equal to $ 4.4 \,\cdot\,10^{-3}$ instead of  $\langle|S^2|\rangle \, =\,7\,\cdot\, 10^{-4}$
The conclusion that we want to make is  that the enhanced Pomeron diagrams can be very
important in obtaining the reliable estimates of the survival probability of the large rapidity gap at the LHC
energy and can lead
to additional suppression of the value of the cross section for diffractive Higgs production.

\TABLE[ht]{
\begin{tabular}{|l|l|l|l|l|l|l|l|l|l|l|}
\hline \hline
 ${\cal Y}$ & 2.5 &   2.75 & 3. & 3.25 & 3.75 & 4.
  & 4.25 & 4.5 &4.75. & 5.  \\
  \hline
  $\langle|S^2|\rangle$  & 0.69 & 0.39 &  0.28 &  0.18 & 0.10 & 0.048 &  0.034 &  0.021 &  0.010 &  0.0044  \\
\hline \hline
  \end{tabular}
  \caption{The values of the survival probabilities versus ${\cal Y}\,=\,\Gamma(1 \to 2)\,Y$
 (the value of  the low energy amplitude $\gamma = 1/\kappa$.)
Calculating $\langle|S^2|\rangle$ \eq{SP44} was used with $\as = 1/4$ and $M_{Higgs}=100GeV$.}
}

\section{Comparison with the  approximate approaches.}
\subsection{Our approach versus the mean field approximation.}
In the mean field approximation we can discuss the single diffractive production
(see \eq{S1} - \eq{S4}) and the value
 of the survival probability. The direct application of the procedure described
in \eq{SP2} - \eq{SP4} for the MFA gives
\bea
&&\gamma \frac{\partial  N\Lb \gamma; Y \Rb}{\partial
\gamma}\,\,=\,\,\frac{\exp\Lb \frac{\cal Y}{2} \Rb}
{\Lb 1\,\,+\,\,\gamma\,\Lb \exp\Lb \frac{\cal Y}{2} \Rb\,\,-\,\,1 \Rb
\Rb^2}\,\,; \label{CP1} \\
&&N\Lb \gamma; Y; \frac{Y}{2} \Rb\,\,=\,\,\frac{\gamma \,\exp\Lb {\cal Y}
\Rb}{\Lb 1\,\,+\,\,\gamma\,\Lb \exp\Lb
\frac{\cal Y}{2} \Rb\,\,-\,\,1 \Rb \Rb^2\,\,+\,\,\gamma\, \exp\Lb \frac{\cal
Y}{2} \Rb\,\Lb
\exp\Lb \frac{\cal Y}{2} \Rb\,\,-\,\,1 \Rb}
\,\,;\label{CP2} \\
&&<|S^2|>\,\,\,\,=\,\,\,\Lb \frac{1}{\Lb 1\,\,+\,\,\gamma\,\Lb \exp\Lb
\frac{\cal Y}{2}\Rb\,\,-
\,\,1 \Rb \Rb^2\,\,+\,\,\gamma\,
 \exp\Lb \frac{\cal Y}{2}\Rb\,\Lb \exp\Lb \frac{\cal Y}{2} \Rb\,\,-\,\,1
\Rb}\Rb^2
\,\,;\label{CP3}
\eea

However, to compare the MFA with our calculations we need to go back from
\eq{N0} and \eq{NIN} to Kovchegov's formulae
\bea
&&N_0\Lb \gamma; Y \Rb\,\,\,=\,\,1\,\,\,-\,\,\,Z\Lb 1\, - \, \gamma; Y \Rb
\,;\label{N01} \\
&&N_{in}\Lb \gamma; Y \Rb\,\,=\,\,1\,\,\,-\,\,\,Z_{in}\Lb 1\, - \, \gamma; Y \Rb
\,; \label{NIN1}
\eea

 The comparison with the MFA is given in \fig{ndmpsi} and \fig{smpsi} for the
diffractive cross section and
survival probability. As it is expected the MFA reproduces the exact result only
for rather large values of the amplitude for
 dipole - target interaction at low energies.
\FIGURE{
\centerline{\epsfig{file= 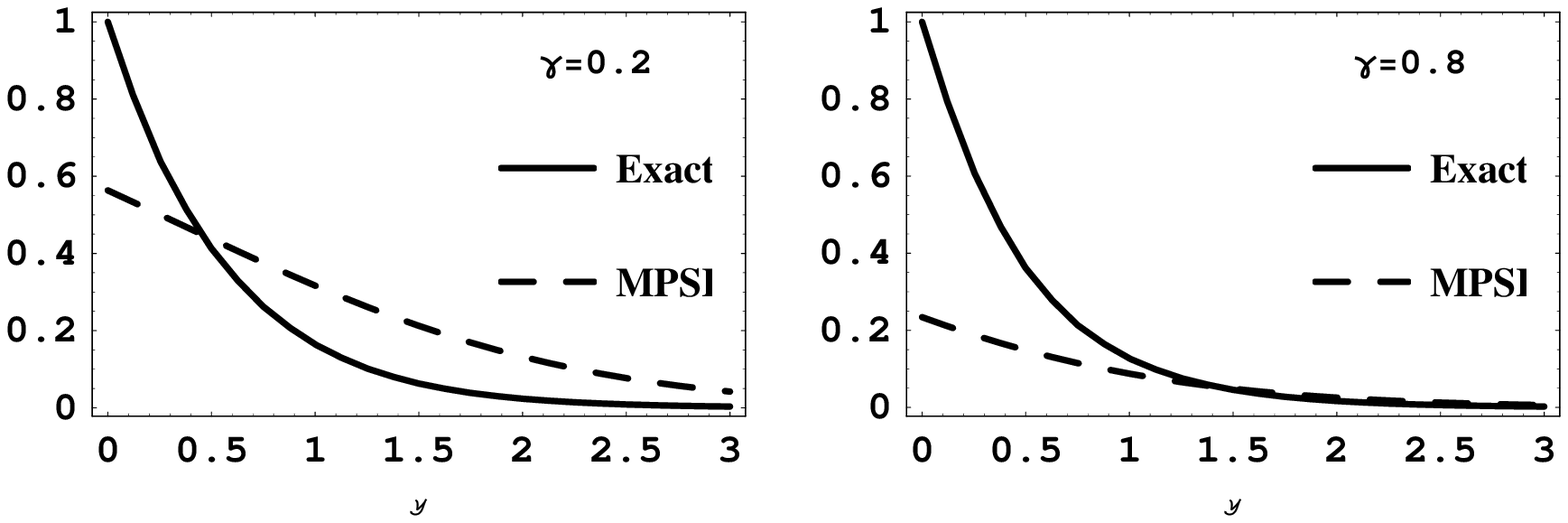,width=170mm}}
\caption{ Survival probability $\langle|S^2\rangle$  versus ${\cal Y} \equiv
\Gamma(1\to 2)\,Y$ at different
values for the low energy amplitude $\gamma$. The solid line is our  calculation
while the dotted line is the MFA approximation. }
\label{smpsi}
}
\FIGURE{
\centerline{\epsfig{file=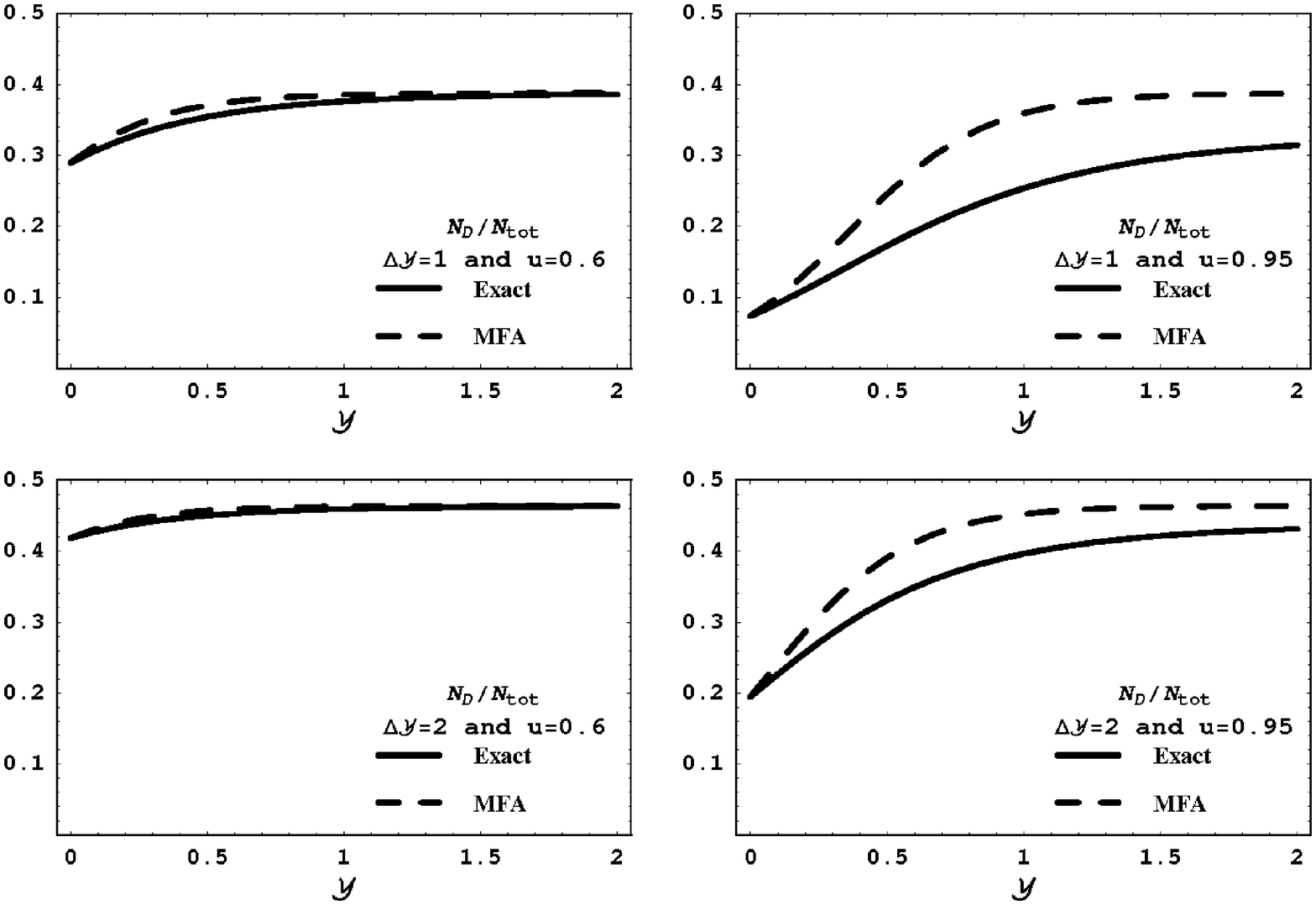,width=170mm}}
\caption{ The ratio $N_D/N_0$ as a function of ${\cal Y}$ for different value of
the amplitude of interaction for
 low energies $\gamma=1 - u$. The solid line is our  calculation while the
dotted line is the MFA approximation. }
\label{ndmpsi}
}

\subsection{Our approach versus Mueller - Patel - Salam - Iancu (MPSI)
approximation.}
The MPSI approximation \cite{MUPA,MS,IM,KOLE}  replaces the mean field one in
the case if the  dipole - target amplitude
 is small at low energies.
The diagrams that contribute to the scattering amplitude in this approximation
are shown in \fig{mpsi}, they all have
 Pomeron
 loops but the size of any loop is larger than $Y/2$.  The general formula of
MPSI approximation can be written in the form
\bea \label{CPS2}
N^{MPSI}_0\Lb \gamma,Y\Rb\,\,\,\,&=&\,\,\,1\,\,\,-\,\,\left\{\exp \Lb
-\,\gamma\,\frac{d}{d \,u_{1 }}\,
\frac{d}{d \,u_{2 }}\Rb\,\,\,
N^{MFA}_0\Lb \frac{Y}{2}\,,u_1 \Rb\,\,N^{MFA}_0\Lb \frac{Y}{2}\,,u_2 \Rb
\,\right\}\mid_{u_{1 } =1; \,u_{2 }=1} \nonumber \\
&=&\,\,1\,\,-\,\,\exp \Lb \frac{1}{\gamma e^{ {\cal Y} }}\Rb\,\frac{1}{\gamma
e^{ {\cal Y} }}\,\,
\Gamma\Lb 0,\frac{1}{\gamma e^{ {\cal Y} }} \Rb
\eea
$\Gamma \Lb 0,x \Rb$ is the incomplete gamma function  (see  formulae {\bf 
8.350 - 8.359} in Ref. \cite{RY}).
We assume that $ e^{ {\cal Y}/2 }\,\,\gg\,\,1$ in \eq{CPS2}.
In \eq{CPS2} we substitute for  $N^{MFA}_0$ the amplitude in the  mean field
approximation but we can use this equation
using any of the amplitude calculated in  \eq{S1} - \eq{S4}.
 For example if one needs to calculate the single diffraction production
\eq{CPS2} has the form
\bea
N^{MPSI}_{SD}\Lb \gamma,Y,Y_0\Rb &=& 1\,\,\,-\,\,\left\{\exp \Lb
-\,\gamma\,\frac{d}{d \,u_{1 }}\,\frac{d}{d \,u_{2 }}\Rb\,\,\,
N^{MFA}_{SD}\Lb \frac{Y}{2}\,,u_1 \Rb\,\,\tilde{N}^{MFA}_0\Lb \frac{Y}{2}\,,u_2 \Rb
\,\right\}\mid_{u_{1 } =1; \,u_{2 }=1} \label{CPS3}\\
&=&\,\,\,\frac{2\,\,e^{ {\cal Y} - {\cal Y}_0}}{ T^2_1}\,\,\left\{
\frac{1}{\gamma \,T_1}\,\,+ \,\,1\,\,-
\,\,\frac{1  +  2\,\gamma \,T_1}{ \Lb \gamma\,T_1\Rb^2}\,\Gamma\Lb 0,
\frac{1}{\gamma \,T_1} \Rb
\,\exp\Lb\frac{1}{\gamma \,T_1}\, \Rb
\right\} \label{CPS4}
\eea
where  
$T_1\,\equiv\,T\Lb Y,Y_0 \Rb\,\,=\,\,2\,\exp \Lb {\cal Y} \Rb \,-\,\exp \Lb 
{\cal Y}/2\,+\, {\cal Y}_0 \Rb$.

In derivation of \eq{CPS3} we use not \eq{S3} but simpler expression, namely \cite{BGLM,DGBK}
\beq \label{CPS41}
\tilde{N}^{MFA}_0(Y,Y_0)\,\,=\,\,\frac{2 (1-u)e^{\cal Y}\,\left( 2 e^{\cal Y}\,-\,e^{ {\cal Y}_0}\,-\,1 
\right)}{\left(
1\,\,+\,\,(1 - u)\,\left( 2\,e^{\cal Y} \,-\,e^{{\cal Y}_0} \,-\,1 \right)\right)^2}
\eeq
This simple formula does not take into account the elastic scattering of the dipole while \eq{S3} does. Therefore, we 
find in \eq{CPS3} the equation for the inelastic diffraction. Practically, such approach could be useful for the 
diffraction in hadron-hadron interaction, however, it is certainly is not good approximation for deep inelastic 
scattering where the elastic scattering also included in the experimental diffractive cross se3ction.

Substituting in \eq{CPS3} $\tilde{N}^{MFA}_D$ given by
\beq \label{CPS42}
\tilde{N}^{MFA}_D\,\,=\,\,2\,(1 - u)e^{\cal Y}\,\left(\frac{1}{1\,\,+\,\,+\,\,(1 - u)\,\left(e^{\cal Y} 
\,-\,\right)} 
\,\,-\,\,\frac{1}{1 \,\,+\,\,(1 -u) \left( 2\,e^{\cal Y}\,\,-\,\,e^{{\cal Y}_0}\,\,-\,\,1 \right)} \right)
\eeq 
we obtain 
\bea \label{CPS43}
&&N^{MPSI}_D(\gamma;Y,Y_0)\,\,= \\
&&2\left\{1 \,-\,\frac{1}{\gamma\,e^{\cal Y}}\,\exp\left(\frac{1}{\gamma\,e^{\cal 
Y}}\right)\,\Gamma\left(0,\frac{1}{\gamma\,e^{\cal Y}}\right) \,\,-\,\,\frac{e^{\cal Y}}{T_1}\,\left( 
1\,\,-\,\,\exp\left(\frac{1}{\gamma\,T_1}\right)\,\frac{1}{\gamma\,T_1}\,\Gamma\left(0,\frac{1}{\gamma\,T_1} 
\right)\right)\right\} \nonumber
\eea

Repeating the same calculation but for double diffraction using \eq{CPS3} but
with substitution $N_0 \to N_{SD}$ we obtain the following
\bea \label{CPS5}
&&N^{MPSI}_{DD}\Lb \gamma,Y,Y_1,Y_2\Rb \,\,=   \\
&&\,\,4\,\,
\frac{e^{{\cal Y} - {\cal Y}_1 - {\cal Y}_2}}{T^2_2} \,\times \,
\left\{\frac{1}{\Lb \gamma\,T_2\Rb^2}\,\,+\,\,\frac{4}{
\gamma\,T_2}\,\,+\,\,1\,\,
-\,\,\frac{1}{\gamma T_2}\,\,\Gamma\Lb 0,\frac{1}{T_2}\Rb\, \times
\exp\Lb  \frac{1}{ \gamma\,T_2 }\Rb
\,\left[\frac{1}{\Lb \gamma\,T_2 \Rb^2}\,+\,\frac{5}{ \gamma\,T_2}\,\,+\,\,
4 \right] \right\}\nonumber
\eea
where $T_2 \,\equiv\,T\Lb Y,Y_1,Y_2 \Rb\,\,=\,\Lb 2\,\exp \Lb \frac{{\cal Y}}{2}
\Rb \,-
\,\exp \Lb {\cal Y}_1 \Rb \Rb\,\,\Lb 2\,\exp \Lb \frac{{\cal Y}}{2} \Rb
\,-\,\exp \Lb {\cal Y}_2 \Rb \Rb$.

 Finally, for the survival probability we obtain in MPSI approach the following
expression
\beq \label{CPS6}
<|S^2|>\,\,=\,\,-\frac{1}{\Lb \gamma e^{{\cal Y}}\Rb^2} \,\,+\,
\frac{1 + \gamma e^{{\cal Y}}}{\Lb \gamma e^{{\cal Y}}\Rb^3}\,\exp \Lb \frac{1}{
\gamma e^{{\cal Y}}}\Rb\,
\Gamma\Lb 0,\frac{1}{\gamma\,e^{{\cal Y}}}\Rb
\eeq
In writing \eq{CPS6} we assume that $e^{{\cal Y}}\,\,\gg\,\,1$.
\FIGURE[h]{
\begin{minipage}{100mm}{
\centerline{\epsfig{file=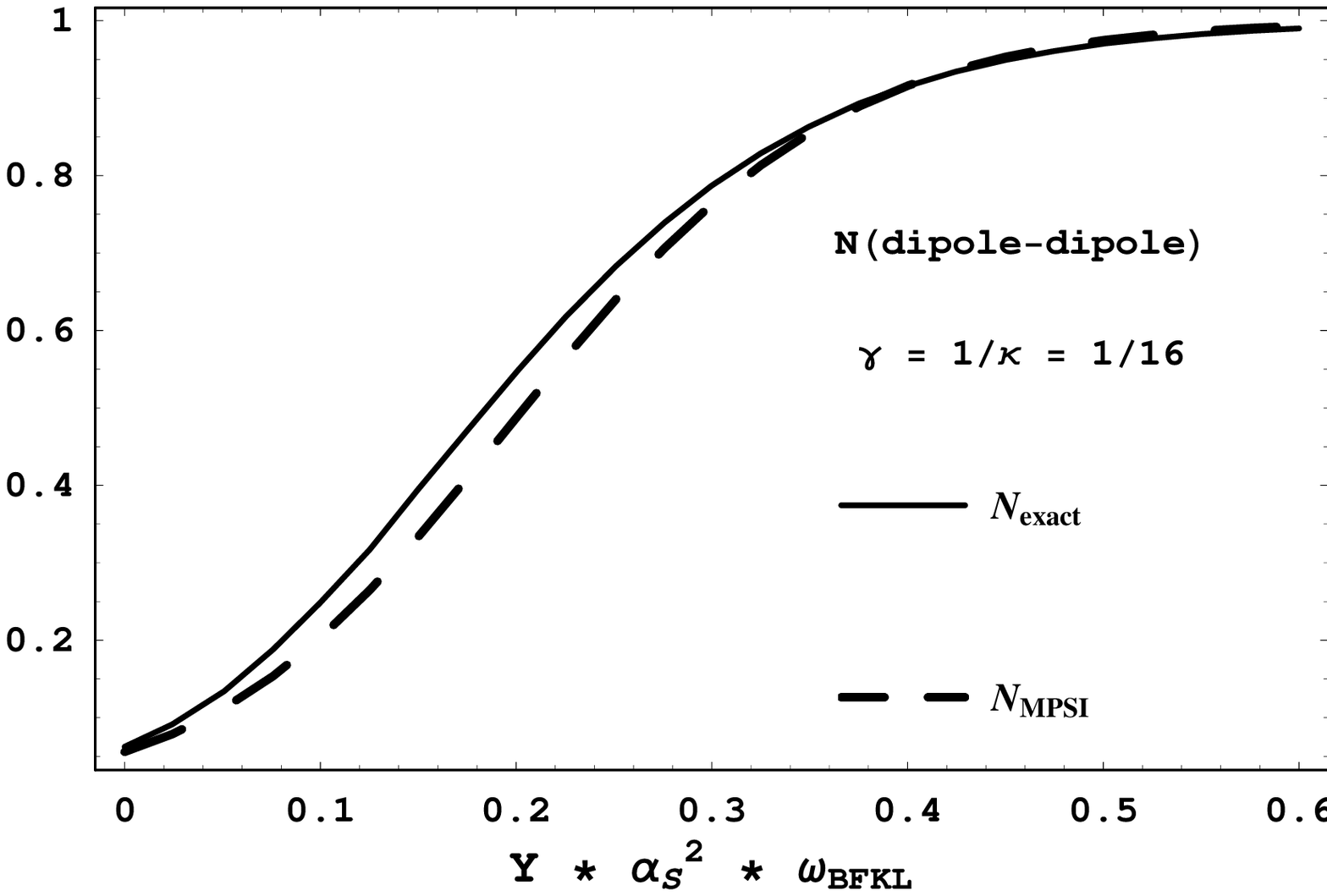,width=100mm}}
}
\end{minipage}
\caption{The comparison of the scattering amplitude for the dipole-dipole
scattering ($N_0$) calculated using \eq{N0}.
The $x$-axis is ${\cal Y} \,=\,\,\omega_{BFKL}\,\as^2\,Y\,\,=\,\,(4\,ln
2\,\bas/\kappa)\,Y$.
The solid
line is the exact solution while the dotted one shows the MPSI result.  }
\label{nmpsi}
}

For a comparison of the exact solution with the MPSI one we need to start from
the dipole-dipole amplitude. In Ref. \cite{KLTM}
we use Kovchegov's formula (see \eq{KAMP}) for calculating the scattering
amplitude. The key idea of this formula is that two or
more dipoles at low energies interact with the target without correlations or in
other words according to the Glauber approach.
 However, for the case of the dipole-dipole
scattering in the  framework of perturbative  QCD approach it was shown (see
Refs. \cite{BART,BRN}) that  two or more dipoles do not
interact with the target dipole at low energy. In this case we have to use
\eq{N0} and \eq{NIN} for calculating our amplitude.
In \fig{nmpsi} one can see the result of these calculation. One can see from
this figure that the MPSI approach describes quite well
the high energy limit of the scattering amplitude but it underestimates the
value of the amplitude at lower energies. This result is
actually quite surprising since the  MPSI approach was proven to be a good
approximation in the limited range of energies, namely,
\beq \label{MPSIR}
\frac{1}{\bas}\,\,\,\,\ll\,\,\,\,Y\,\,\,\,\ll \,\,
\frac{1}{\bas}\,\ln\Lb \frac{N^2_c}{\bas^2} \Rb
\eeq

 For our choice of $\kappa  = 16$ and $\as = 1/4$ in \fig{nmpsi} ( considering
$\omega_{BFKL} = 4\ln2\,\bas$)  \eq{MPSIR} gives
$0.044\, < \, {\cal Y} = ( \omega_{BFKL}\,/\kappa ) \,Y\,<\,0.2$. One can see
that in this range the MPSI approximation is not able to describe the exact solution.

In \fig{ndmpsi1} we see that even for the single diffractive production the 
MPSI approximation describes the exact result quite well.

\section{Conclusions}
In this paper we obtain two main results.
The first result of  this paper is the set of formulae in the framework of the
BFKL Pomeron calculus in
 zero transverse dimensions, that
describe the diffractive dissociation processes in the limited range of produced
mass given by \eq{MR} - \eq{MR2}. The comparison with the approximation methods,
that there exists on the market,
shows that these methods are  not able to describe the diffractive production in
the entire kinematic region.
However, the mean field approximation gives a very good description if the low
energy amplitude of interaction
with the target is rather
large and if the two or more dipoles interact with the target without
correlations according
 to the Glauber formulae.

This great surprise stems from the fact that the MSPI approximation is able to
describe the high energy limit of
the amplitudes for elastic
and diffractive scattering while theoretically this approach can be quaranteed
only in the limited kinematic
 range given by \eq{MPSIR}.

The surprise comes from two observations: first, the MSPI result does not 
lead to the decreasing cross section in spite of the fact that onle triple 
Pomeron interaction has been taken into account; and, second, for the 
total cross section  the MSPI estimates are very close to the exact 
solution (see \fig{nmpsi}).

Having in mind these two observations we believe that the MSPI 
approximation can be used  as the simple first approximation instead of 
the MFA for the case of dipoloe-dipole scattering. However4, we would like 
to stress that for the value of the survival probability the MPSI 
approximation gives leads to higher probability of survival of the large 
rapidity gaps than the exact answer.

The second result is the formula for the survival probability for the central
diffractive (see \eq{MR3}
 production of Higgs meson (di-jets
and so on ). This formula is exact and it is valid for any value of energy. The
estimates using this formula gives
the value for the survival
probability which is much smaller (in 10 - 100 times) than the estimates of all
approximation methods
that have been used before. We hope that this result will stimulate a more
detailed study of the survival probability in QCD
and we consider the first papers on this subject (see Refs.\cite{BBKM,JM} as
only a beginning for such kinds of studies.

In general,  we hope that this paper, devoted to a model: the BFKL Pomeron
calculus in zero transverse dimension, will be useful in searching
the high energy asymptotic behaviour in the general case of QCD and/or in the
case of the BFKL Pomeron calculus in two  transverse
dimensions.

\FIGURE[h]{
\begin{minipage}{150mm}{\
\centerline{\epsfig{file=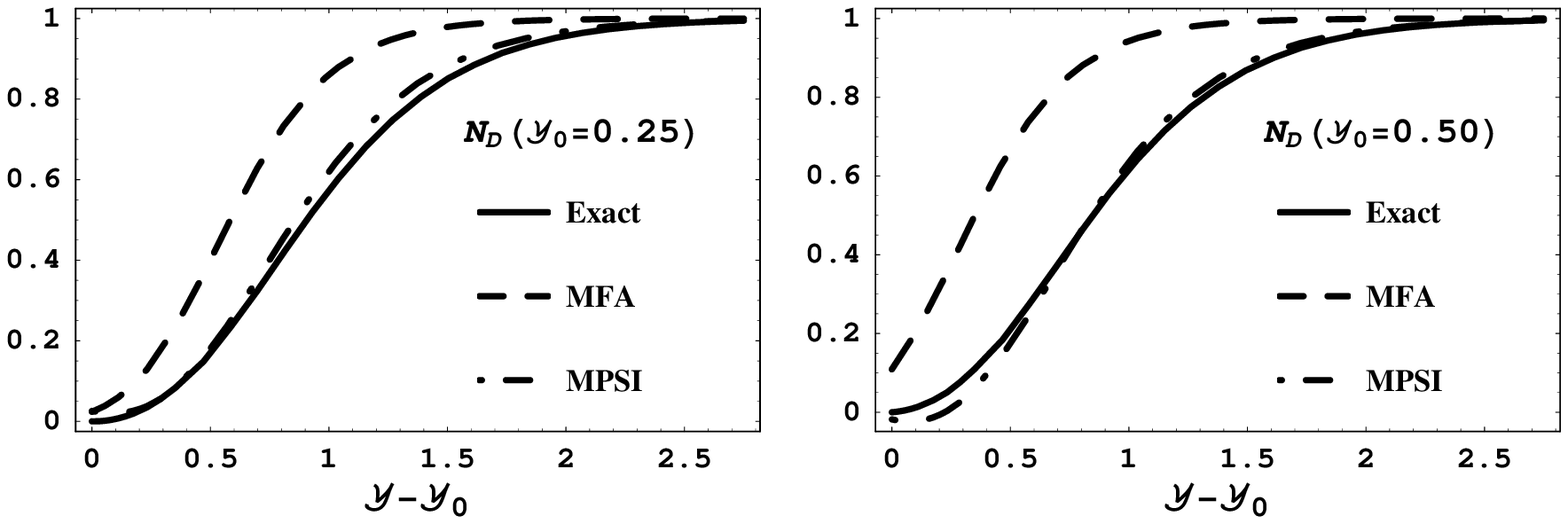,width=150mm}}
}
\end{minipage}
\caption{The comparison of the   cross section of diffractive production
with mass $\ln M^2 \geq\,{\cal Y}_0/(\omega_{BFKL} \,\as^2)$   for the
dipole-dipole scattering.
The $x$-axis is ${\cal Y}- {\cal Y}_0
\,=\,\,\omega_{BFKL}\,\as^2\,(Y - Y_0)\,=\,\,(4\,ln
2\,\bas/\kappa)\,(Y - Y_0)$.
The solid
line is the exact solution while the dotted one shows the MFA result and
dashed-dotted line shows the MPSI calculation.  }
\label{ndmpsi1}
}

We would like to draw attention to the fact that in this paqper we sum 
mostly the so called emhanced diagrams (see \eq{N0} and \eq{NIN} and 
\fig{sdgen},\fig{ddgen} and \fig{hp}). These diagrams cannot give the 
correct answer since they did not reproduce the elastic cross section. The 
elastic cross section is equal to zero in this approach. However, 
accordingly to the general idea of the Pomeron calculus the sum of these 
diagrams gives us the new resulting Green's function ($G_P$ of the Pomeron 
and we 
should try to build a new theory of the interacting Pomeron based on this 
function.
It is easy to see that at $Y \to \infty$ 
\beq \label{GP}
G_P(Y,u)\,\,\to\,\,\frac{1}{1 - e^{- \kappa}}\, \,>\,\,1
\eeq
Therefore, the exchange of the resulting Pomeron leads (i)  to the 
amplitude 
larger than unity but (ii) this amplitude is very close to the unitarity 
bound ($G_P =1$). Therefore, we have a kind of black disc behaviour 
without the elastic cross section. We would lioke to recall that in the 
black disc limit $\sigma_{el} = 1/2 \sigma_{tot}$.

The fast decrease of the single diffractive production (see \fig{nsd}) 
shows that  the effective triple Pomeron interaction vahishes at large 
$Y$. 
In other words,  we can consider $G_P$ given by \eq{GP} as the exact 
solution of the multi interacting Pomeron system. On the other hand, the 
sum of enhanced diagrams gives the model for the two paricles irreducible 
diagrams in $s$-channel. To sum all diagrams we need to take into account 
also all two particle reducible diagrams. This sum is given by the 
eikonal (Glauber)  formula, namely,
\beq \label{AGP}
A_{asymp} \,\,\to\,\,1\,\,-\,\,e^{-G_P}\,\,=\,\,1 \,\,-\,\,\exp\left(- 
\frac{1}{ 1 - e^{- \kappa}}\ \right)\,\,<\,\,1
\eeq
Therefore we obtain a grey disc limit. The eikonalization gives a simple 
solution for all observables calculated here. However, \eq{AGP}  
deserves
a more careful consideration, as was noticed in Ref. \cite{BMMSX}, before 
being used.  \eq{AGP}  gives the behaviour of the scattering amplitude at 
low energies but at $u \to 0$ ($\gamma \to 1)$. Therefore, solving the 
Pomeron calculus we reduce the problem of high energy scattering to the 
scattering at low energies while the last step: to find the scattering 
amplitude at $u \to 1$ remains open and, perhaps, eikonalization is one of 
possible solutions of this problem.

\section*{Acknowledgments:}
We want to thank Asher Gotsman, Edmond Iancu, Uri Maor  and Al Mueller
for very useful discussions on the subject of this paper.

This research was supported in part  by the Israel Science Foundation,
founded by the Israeli Academy of Science and Humanities and by BSF grant \#
20004019.

\end{document}